\documentclass[12pt]{article}
\pdfoutput=1
\usepackage{jheppub}
\usepackage{tabu}
\usepackage[vcentermath]{youngtab}
\usepackage[usenames,dvipsnames,table]{xcolor}
\usepackage{graphicx,amsmath,amssymb,amsthm,multirow,array,bm,esint}
\usepackage[mathscr]{eucal}
\usepackage{epsf,amsfonts}
\usepackage{slashed}
\usepackage[numbers,sort&compress]{natbib}
\usepackage{minibox}
\usepackage{rotating}
\usepackage{pdflscape}
\usepackage{array,tikz-cd}
\usepackage{subfigure}
\usepackage{pgfplots}
\pgfplotsset{compat=1.15}
\usepackage{mathrsfs}
\usetikzlibrary{arrows}
\usetikzlibrary{backgrounds}
\usepackage{slashed}
\usepackage{tikz}
\usetikzlibrary{decorations.markings}
\tikzset{middlearrow/.style={
        decoration={markings,
            mark= at position 0.5 with {\arrow{#1}} ,
        },
        postaction={decorate}
    }
}

\usepackage{import}
\usepackage{pdfpages}
\usepackage{transparent}

\definecolor{rust}{rgb}{0.8,0.2,0.2}

\definecolor{rust}{rgb}{0.8,0.2,0.2}

\definecolor{labelkey}{rgb}{0.4,0.4,0.4}

\title{ Reduced Half-sided Translations and Islands}

\author{Krishna Jalan,}
\author{Roji Pius}

\affiliation[a]{
The Institute of Mathematical Sciences, IV Cross Road, C.I.T. Campus, Taramani, Chennai, India 600113}

\affiliation[b]{
Homi Bhabha National Institute, Training School Complex, Anushakti Nagar, Mumbai, India 400094}

\emailAdd{krishnajalan@imsc.res.in}
\emailAdd{rojipius@imsc.res.in}

\abstract{ 
 An AdS$_2$ black hole in equilibrium with a finite temperature non-gravitating bath comes with a Hawking-like information paradox. The resolution of this paradox requires introducing an island for the bath after the Page time. Since the island region contains the black hole interior, the consistency of the island paradigm demands the failure of any interior reconstruction proposal that uses only the operators in the AdS$_2$ region outside the black hole horizon after the Page time. In this paper, we investigate the consistency of the island paradigm using the black hole interior reconstruction proposal due to Leutheusser and Liu. They argued that the operators in the interior of a black hole can be reconstructed by the half-sided translations of the operators outside the black hole horizon. However, in order to illustrate the island paradigm we need to modify the Leutheusser-Liu proposal by introducing the notion of the reduced half-sided translations. The reduced half-sided translations, unlike the half-sided translations, have non-trivial action only on the operators in the AdS$_2$ black hole region. As a result, the reconstructed interior operators are expressed only using the operators in the AdS$_2$ region outside the black hole horizon. We demonstrate the consistency of the island paradigm by showing that the reduced half-sided translation, which successfully reconstruct the operators in the black hole interior before the Page time, fails to reconstruct the interior operators after the Page time. }

\begin{document}
\maketitle
 
\section{Introduction}
\label{sec:intro}
Resolving the black hole information paradox \cite{Hawking:1976ra,Hawking:1975vcx} is one of the biggest challenges in semiclassical gravity. A consistent resolution of this puzzle involves pinpointing the right semi-classical gravity computation for the entropy of the Hawking radiation that reproduces the Page curve \cite{Page:1993df}. Progress in recent years suggest that the paradox can be resolved by identifying the entropy of the Hawking radiation with the von Neumann entropy of the state of the radiation \cite{Almheiri:2020cfm}. The computation of the von Neumann entropy of the Hawking radiation can be performed in a two dimensional toy model of an evaporating black hole obtained by attaching a non-gravitating bath to the boundary of an AdS$_2$ black hole \cite{Almheiri:2019qdq}. The Euclidean path integral computation of the radiation entropy demonstrated that after the Page time the leading contribution to the entropy comes from a Euclidean wormhole saddle \cite{Goto:2020wnk}. The net effect of the wormhole contribution is the inclusion of an island region which is far away from the non-gravitating bath in the radiation entropy computation \cite{Almheiri:2019hni}. Since the island region contains the black hole interior, it is expected that the reconstruction of the operators inside the black hole horizon is not possible using the operators living in the AdS$_2$ region outside the horizon after the Page time.
\par

 Recently, Leutheusser and Liu argued that the operators inside the horizon of a black hole can be reconstructed by the half-sided translations of the operators outside the horizon \cite{Leutheusser:2021qhd,Leutheusser:2021frk}\footnote{Similar reconstruction is possible by following the proposal of Papadodimas and Raju \cite{Papadodimas:2012aq, Papadodimas:2013jku}.}. The half-sided translations are transformations generated by the difference of the modular Hamiltonians associated with an algebra of observables $\mathcal{M}$ and its subalgebra $\mathcal{N}$ \cite{Borchers:2000pv}. For an AdS$_2$ eternal black hole, we can choose $\mathcal{M}$ to be the algebra of observables inside a wedge region $W_{\mathcal{M}}$ in the right side of the horizon and $\mathcal{N}$ to be the algebra of observables inside another wedge region $W_{\mathcal{N}}$, which is completely inside $W_{\mathcal{M}}$. The resulting half-sided translations can translate operators outside the horizon to the black hole interior. This can be interpreted as the reconstruction of an interior operator in terms of the exterior operators. \par

A more non-trivial example is that of an eternal AdS$_2$ black hole in equilibrium with a finite temperature bath. In this case, the continuous exchange of Hawking quanta emitted by black hole and the bath quanta builds an ever-growing entanglement between the black hole and the bath\footnote{The classical geometry of the eternal black hole is unaffected by this exchange due to the vanishing net flow of stress tensor across the cut out boundary of the AdS$_2$ region.}. This gives rise to a linear growth of von Neumann entropy for the black hole with respect to the boundary time, which eventually grows beyond the thermodynamic entropy of the black hole. The resulting information paradox can be resolved by transferring the ownership of the interior of the black hole from the left and right regions of the black hole to the bath after the Page time \cite{Almheiri:2019yqk}. Hence according to the entanglement wedge reconstruction proposal \cite{Dong:2016eik}, the black hole interior is not expected to be reconstructable using the exterior operators in the AdS$_2$ regions after the Page time. The goal of this paper is to demonstrate this fact by using the half-sided translations. However, the half-sided translations constructed using the algebras of observables defined in the gravitational region also depends on the operators in the non-gravitational bath. This is because the half-sided translations not only depend on the algebras of observables but also on their commutants\footnote{Commutant of an algebra of observables is the set of all observables that commute with every element of it.}. As a result, the interior operators reconstructed using the half-sided translations are not solely expressible in terms of the exterior operators in the AdS$_2$ region. \par

In this paper, we circumvent this by introducing the notion of the ``reduced half-sided translations''. These are constructed using the algebras of observables and their {\it relative commutants} \cite{Borchers:2000pv} within the gravitational region. We demonstrate the consistency of the island paradigm by showing that the reduced half-sided translations which successfully reconstruct the operators in the black hole interior before the Page time become trivial and hence fail to reconstruct the interior operators after the Page time.\par

The paper is organised as follows. In section \ref{HST}, we briefly review the construction of the half-sided translations and explain how it generates an in-falling time evolution across a Rindler horizon. In section \ref{IRHST}, we describe the island paradigm and the need of introducing the notion of the reduced half-sided translations. The action of the reduced half-sided translation before the Page time is explained in section \ref{ARHSTBP}.  By assuming the existence of the reduced half-sided translation in quantum JT gravity,  in section \ref{RHSTJT} we describe how the reduced half-sided translation can be understood as its saddle point approximation. We also argue the need of change of saddle after the Page time due to the unbound growth of free energy of the saddle that described the reduced half-sided translation before the Page time.  In section \ref{RHSTAPT}, we describe the construction of a wormhole saddle having constant free energy after the Page time and identified the corresponding modified reduced half-sided translation. We show that this modified reduced half-sided translation after the Page time successfully spread an operator in the left bath to black hole interior.  We conclude in section \ref{disc} by discussing some of the interesting future directions.

\section{The half-sided translations and the in-falling evolution}\label{HST}
The main ingredient of the Leutheusser-Liu proposal for reconstructing the operators in the black hole interior is the half-sided translations. In this section, we shall briefly discuss the definition of half-sided translations and describe the action of half-sided translation on a primary field in an arbitrary two dimensional CFT defined in flat spacetime. 

\subsection{The half-sided translations}
The half-sided translations are transformations defined with respect to a von Neumann algebra $\mathcal{M}$, and a subalgebra $\mathcal{N}$ having the same cyclic and separating vector $|\Omega\rangle$. The separating property of $|\Omega\rangle$ implies that it is not annihilated by any of the operators in $\mathcal{M}$, and the cyclic property implies that it is possible to construct a dense subspace of the Hilbert space associated with $\mathcal{M}$ by acting with the elements of $\mathcal{M}$ on $|\Omega\rangle$. The half-sided translations form a continuous one-parametric unitary group. An element of this group is denoted as $U(s)$, where $s$ is the continuous real parameter that parametrises the abelian group. Each element of this group has the property that it keeps the cyclic separating state invariant $$\quad U(s)|\Omega\rangle =|\Omega\rangle \qquad \forall s\in \mathbb{R}.$$ Another important property is that the action of $U(s)$ for $s\leq 0$ keeps the operators within the algebra $\mathcal{M}$ $$ U^{\dagger}(s)\mathcal{M}U(s)\subset \mathcal{M} \qquad \forall s\leq 0.$$ For $s$ equal to minus one we obtain the subalgebra $\mathcal{N}$ $$\mathcal{N}= U(-1)^{\dagger}\mathcal{M}U(-1).$$ However, this structure exists only if $\mathcal{M}$ is a type III$_1$ von Neumann algebra \cite{borchers1998half}. \par

Interestingly $U(s)$ can be constructed using the modular operators $\Delta_{\mathcal{M}}$ and $\Delta_{\mathcal{N}}$ associated with the algebras $\mathcal{M}$ and $\mathcal{N}$ respectively \cite{Borchers:2000pv} as below
\begin{equation}
\label{UsModop}
U(s)=\Delta_{\mathcal{M}}^{-\text{i} t}\Delta_{\mathcal{N}}^{\text{i} t}, \quad \quad s=e^{-2\pi t}-1 \quad \forall~ t\in \mathbb{R}.
\end{equation}
The modular operator $\Delta_{\mathcal{M}}$ preserves the operator algebra $\mathcal{M}$ and keeps the cyclic and the separating vector invariant. Moreover, it acts both on $\mathcal{M}$ and its commutant $\mathcal{M}'$. In general, $\Delta_{\mathcal{M}}$ can not be factorised into two operators $\bar\Delta_{\mathcal{M}}$ and $\hat\Delta_{\mathcal{M}}$ which acts only on $\mathcal{M}$ and $\mathcal{M}'$ respectively. It can be expressed as the exponential of the modular Hamiltonian $\mathcal{K}_{\mathcal{M}}$ of the algebra $\mathcal{M}$, $\Delta_{\mathcal{M}}=e^{-\mathcal{K}_{\mathcal{M}}}$. The modular Hamiltonians of $\mathcal{M}$ and $\mathcal{N}$ satisfies the following commutation relation \cite{Leutheusser:2021frk}
\begin{equation}\label{modMNalgebra}
\left[\mathcal{K}_{\mathcal{M}},\mathcal{K}_{\mathcal{N}}\right]=-2\pi \text{i}\left(\mathcal{K}_{\mathcal{M}}-\mathcal{K}_{\mathcal{N}}\right).
\end{equation}
Using (\ref{UsModop}), (\ref{modMNalgebra}) and the Baker-Campbell-Hausdorff formula we can then express $U(s)$ as follows
\begin{equation}
\label{UsModopgen}
U(s)=e^{-\text{i}sG},
\end{equation}
where the generator $G$ of the group of the half-sided translations is given by $$G=\frac{1}{2\pi}(\mathcal{K}_{\mathcal{M}}-\mathcal{K}_{\mathcal{N}}).$$

\subsection{The half-sided translation in two dimensional Minkowski spacetime}
Let us discuss the action of half-sided translations on a local operator in an arbitrary two dimensional conformal field theory having central charge $c$ defined in Minkowski spacetime having light cone coordinates $(y^+,y^-)$. Choose $\mathcal{M}$ to be the algebra of observables in a causal diamond $\Diamond_{{M}}$. Denote by $\Diamond_{{M}}$ the domain of dependence of the region $M$ given by $y^-<0$ on the equal time Cauchy slice $\Sigma_0$. It can be identified with the right Rindler wedge of the Minkowski spacetime. The subalgebra $\mathcal{N}$ is chosen to be the operator algebra in another causal diamond $\Diamond_{N}$ inside $\Diamond_{M}$ as shown in the figure \ref{fig:MinkrindlerMN}. The causal diamond $\Diamond_{N}$ can be obtained by translating all the points in $\Diamond_{M}$ along the $y^-$ direction by an amount $y^-_{N}$. It is the domain of dependence of the region $N$ given by $y^-<y^-_N$ on the Cauchy slice $\Sigma_{t_N}$, where $ t_N=y^+_N-y^-_N$. Here $(y^+_{N},y^-_N)$ are the coordinates of the point $q_N$, the left tip of the diamond $\Diamond_{N}$. The cyclic and separating state is chosen to be the Minkowski vacuum.\par

\begin{figure}
\centering
\begin{tikzpicture}[scale=.95]
 \draw [thick,](1 mm, 10 pt) (1.5,2.5)--(-1,.25) node[pos=.5,sloped, below] {$y^-=-\infty$} ;
 \draw [thick,](1 mm, 10 pt) (-6,4.75)--(-3.5,7) ;
 \draw [thick,](1 mm, 10 pt) (-1,4.75)--(-3.5,7) ;
 \draw [thick,](1 mm, 10 pt) (-6,.25)--(-3.5,-1.9) ;
 \draw [thick,](1 mm, 10 pt) (-1,.25)--(-3.5,-1.9) ;
 \draw [thick](1 mm, 10 pt) (1.5,2.5)--(-1,4.75)node[pos=1,sloped, above] {$y^+=\infty$} ;
 \draw [thick,](1 mm, 10 pt) (-8.5,2.5)--(-6,.25) node[pos=1,sloped, below] {$y^+=-\infty$} ;
 \draw [thick](1 mm, 10 pt) (-8.5,2.5)--(-6,4.75)node[pos=1,sloped, above] {$y^-=\infty$};
 \draw [thick](1 mm, 10 pt) (-8.5,2.5)--(-6,4.75)node[pos=1,sloped, above] {$y^-=\infty$};
 \fill[fill=blue!20] (-.5,4.3)--(-3,2.05)--(-1,.25)--(1.5,2.5);
 \fill[fill=red!20] (-6,4.75)--(-3,2.05)--(-5.425,-.22)--(-8.5,2.5);
\draw(-3.7,4.45) node[above] {$\mathcal{O}(p')$};
\draw(-3.7,4.45) node [circle,fill,inner sep=1pt]{};
\draw (-0.6,1.4) node[right] {$\mathcal{O}(p)$};
\draw(-0.6,1.4)node [circle,fill,inner sep=1pt]{};
 \draw [violet, very thick] plot[smooth]coordinates {(-8.5,2.5)(-6,3.75) (-3.5,4.5) (-1,3.75)(1.5,2.5) } ;
 \draw [violet, very thick] plot[smooth]coordinates {(-8.5,2.5)(-6,1.25) (-3.5,.5) (-1,1.25)(1.5,2.5) } ;
 \draw [color=red,thick,middlearrow={<}](1 mm, 10 pt) (-1.7,2.45)--(-1.1,1.9) ;
 \draw [color=red,thick,dashed](1 mm, 10 pt) (-3.7,4.45)--(-0.6,1.4) ;
 \draw [ blue,thick,dotted,middlearrow={<}] plot [smooth] coordinates { (-1.7,2.45) (-1.55,1.75)(-1.35,1.35) (-1.05,1)};
 \draw [ black,thick,dotted,->] plot [smooth] coordinates { (-1.1,1.9) (-1.1,1.6)(-1.095,1.4) (-1.05,1)};
\draw (-.5,2.5) node {$\mathcal{N}$};
\draw (-6,2.5) node {$\mathcal{N}'$};
\draw (-3.5,2.5) node [circle,fill,inner sep=1.5pt]{};
\draw (-3.5,2.8) node {$q_M$};
\draw (-3,2.05) node [circle,fill,inner sep=1.5pt]{};
\draw (-3,1.75) node {$q_N$};
\draw [color=black!50!green,thick](1 mm, 10 pt) (-6,4.75)--(-1,.25) ;
\draw [color=black!50!green,thick](1 mm, 10 pt) (-6,.25)--(-1,4.75);
\end{tikzpicture}
\caption{The half-sided translation is defined using the von Neumann algebra $\mathcal{M}$ restricted to the right Rindler wedge and the subalgebra $\mathcal{N}$ restricted to the blue shaded region of the two dimensional Minkowski spacetime. It translates a field $\mathcal{O}$ from point $p$ to $p'$ along a null ray towards the Rindler horizon for $s>0$. This half-sided translation is generated by the difference between two boost generators $\mathcal{K_M}$ and $\mathcal{K_N}$, the modular Hamiltonians associated with the algebra $\mathcal{M}$ and $\mathcal{N}$ respectively. The blue dotted arrow shows the trajectory of $\mathcal{K_M}$, the boost that fixes the point $q_{{M}}$, and the gray dotted arrow shows the trajectory of $\mathcal{K_N}$, the boost that fixes the point $q_{{N}}$. The difference $\mathcal{K_M}-\mathcal{K_N}$ generates translation along the null direction represented by the red line.} \label{fig:MinkrindlerMN}
\end{figure}

 The modular operator associated with an algebra defined in a wedge region, unlike in the general case, factorises into a piece that acts on the wedge algebra and another piece that acts only on its commutant. Since $\mathcal{M}$ is a wedge algebra, the modular operator takes the form 
\begin{equation}\label{DMrho}
\Delta_{\mathcal{M}}=\Delta_{\mathcal{M}}^L \Delta_{\mathcal{M}}^R,
\end{equation}
where $ \Delta_{\mathcal{M}}^R$ acts only on the operators in $\mathcal{M}$ and $ \Delta_{\mathcal{M}}^L$ acts only on the operators in $\mathcal{M}'$. The operator $ \Delta_{\mathcal{M}}^R$ and $ {\Delta^L_{\mathcal{M}}}^{-1}$ can be understood as the reduced density matrices $\rho_{{M}}$ and $\rho_{{M}'}$ associated with the regions ${M}$ and ${M}'$, the complement region of $M$ in $\Sigma_0$ respectively. Therefore, the modular operator $\Delta_{\mathcal{M}}$ is given by 
\begin{equation}
\label{modMrhoe}
\Delta_{\mathcal{M}}=\rho^{-1}_{{M}'} \rho_{{M}}. 
\end{equation}
 Likewise, the modular operator $\Delta_{\mathcal{N}}$ is given by
 \begin{equation}
\label{modNrhoe}
\Delta_{\mathcal{N}}= \rho^{-1}_{{N}'}\rho_{{N}}.
\end{equation}
where $\rho_{{N}}$ and $\rho_{{N}'}$ are the reduced density matrices associated with the regions ${N}$ and ${N}'$, the complement region of $N$ in $\Sigma_{t_N}$ respectively. Therefore, the half-sided translation operator $U(s)$ can be expressed as follows
\begin{equation}
\label{UsMink}
U(s)=\rho_{{M}'}^{\text{i} t}\rho^{-\text{i} t}_{{M}}\rho^{-\text{i} t}_{{N}'}\rho_{{N}}^{\text{i} t} \quad \quad s=e^{-2\pi t}-1 \quad \forall ~t\in \mathbb{R}.
\end{equation}

The density matrix $\rho_{{M}}$ can be prepared by performing the CFT path integral over the complex plane, obtained by the Euclidean continuation of the two dimensional Minkowski spacetime, with a cut from $q_{{M}}$ to $i^0_R$. Similarly, the density matrix $\rho_{{M}'}$ is obtained by performing the CFT path integral over the plane with a cut from point $q_{{M}}$ to $i^0_L$. Hence, the path integral representation of $\rho_{{M}}^{\text{i} t}$ is given by the Euclidean CFT path integral in a wedge having clockwise angle $\theta=\text{i}t$ \cite{Witten:2018zxz}. Interestingly, the path integral over a wedge can be replaced by the following conical twist field operator 
\begin{equation}
\label{contwist}
\mathcal{T}_{\theta}(y_{{M}},\bar y_{{M}})\propto e^{\theta\int_{{C}_M}(y-y_{{{M}}})T(y)dy+\text{c.c.}},
\end{equation}
where $T(z)$ is the holomorphic stress-tensor, ${C}_M$ is a curve that connects the points $q_M$ and $i^0$, and $(y_{{M}},\bar y_{{M}})$ are the complexified coordinates of the point $q_{{M}}$ \cite{Castro-Alvaredo:2017wzf}. The job of the proportionality factor is to make the conical twist field a spinless primary field\footnote{The scaling dimension of this primary field is $\frac{c}{12}\left(\frac{2\pi+\theta}{2\pi}-\frac{2\pi}{2\pi+\theta}\right)$, where $c$ is the central charge of the conformal field theory.}. It does not play a significant role in our discussion, more details can be found in \cite{Castro-Alvaredo:2017wzf}. Hence, the modular Hamiltonians $\mathcal{K_M}$ and $\mathcal{K_N}$ are given by
\begin{align}
\mathcal{K}_{\mathcal{M}}=\int_{{C}_M}(y-y_M)T(y)dy+\text{c.c.} \qquad \mathcal{K}_{\mathcal{N}}=\int_{{C}_N}(y-y_N)T(y)dy+\text{c.c.}.
\end{align}
 Substituting this in \eqref{UsModopgen} gives $U(s)$ as follows
 \begin{equation} \label{UsMinkCtwist}
U(s)= e^{-\frac{\text{i}s}{2\pi} \int_{{C}}(y_M-y_N)T(y)dy + c.c.} \quad \quad s=e^{2\pi \text{i}\theta}-1,
\end{equation}
where $C$ is a contour connecting $q_M, q_N, i^0$.

Now we can find the action of $U(s)$ on the operator $\mathcal{O}(y_p,\bar y_p)$. The coordinates $(y_p,\bar y_p)$ are the Euclidean continuation of the coordinates $(y_p^+,y_p^-)$ of the point $p$ in the right Rindler wedge where the operator is kept. The resulting operator $\mathcal{O}_{ s}(y_p,\bar y_P)$ is given by 
 \begin{align}
\label{Utildesplane}
\mathcal{O}_{s}(y_p,\bar y_p)&=U^{\dagger}( s)\mathcal{O}(y_p,\bar y_p)U( s)\nonumber\\
&=e^{-\frac{\text{i}s}{2\pi} \oint_{{C}_p}(y_M-y_N)T(z)dz + c.c.}\mathcal{O}(y_p,\bar y_p),\nonumber\\
&=\mathcal{O}(y_p+ s (y_M-y_N),\bar y_p+ s (\bar y_M-\bar y_N)).
\end{align}
 Here ${C}_p$ is a counter-clockwise contour enclosing the point $p$. Since $y^+_M=y^+_M=0$, the resulting half-sided translation for $s>0$ is an in-falling evolution as shown in \cite{Leutheusser:2021frk}. 

\section{The island paradigm and the reduced half-sided translations }\label{IRHST}
In the previous section, we showed that operators outside the Rindler horizon of a two dimensional Minkowski space can be taken to the interior by the action of half-sided translations. Therefore, the degrees of freedom inside the horizon can be expressed using the degrees of freedom outside the horizon via the half-sided translations. In this section, we will apply this reconstruction procedure to demonstrate the island paradigm. \par
 
 Consider an eternal AdS$_2$ black hole in thermal equilibrium with a finite temperature bath. There is a continuous exchange of Hawking quanta emitted by black hole and the bath which guarantees vanishing net flow of stress tensor across the cut-out boundary of the AdS$_2$ region. As a result, the classical geometry of the eternal black hole remains intact. However, the uninterrupted exchange of quanta builds an ever-growing entanglement between the black hole and the bath. This gives rise to a linear growth of von Neumann entropy for the black hole with respect to the boundary time. Eventually, the entropy of the black hole grows beyond the allowed limit, thermodynamic entropy of the black hole. \par
 
 The resolution of the resulting information paradox requires transferring the ownership of the interior of the black hole from the left and right region of the black hole to the bath, after the Page time \cite{Almheiri:2019yqk}. Hence, the black hole interior is not expected to be reconstructable using the operators in the AdS$_2$ regions outside the left and the right horizons of the black hole after the Page time. The aim of this section is to demonstrate this fact using half-sided translation. We attempt to show that after the Page time, the action of half-sided translations on a local operator in the gravitational region outside the black hole horizon cease to translate a field in the gravitational region outside the event horizon to the black hole interior. \par
 
 \subsection{Eternal black hole in equilibrium with a finite temperature bath}
 \begin{figure}
\centering
\begin{tikzpicture}[scale=1, every node/.style={scale=.85}]
 \draw [red, very thick,dotted] plot [smooth] coordinates {(-6,4.75) (-3.5,3.5) (-1,4.75)} ;
 \draw [thick,](1 mm, 10 pt) (-3.5,7)--(1.5,2.5)node[pos=.5,sloped, above] {$w^+=\infty,~ y_R^+=\infty$} ;
 \draw [thick,](1 mm, 10 pt) (-3.5,-2)--(1.5,2.5) node[pos=.5,sloped, below] {$w^-=-\infty, ~ y_R^-=-\infty$} ;
 \draw [red, very thick,dashed] plot[smooth]coordinates {(-6,0.25) (-3.5,1.5) (-1,0.25)} ;
 \draw [thick,](1 mm, 10 pt) (-3.5,7)--(-8.5,2.5) node[pos=.5,sloped, above] {$y_L^-=\infty,~w^-=\infty$} ;
 \draw [thick,](1 mm, 10 pt) (-3.5,-2)--(-8.5,2.5) node[pos=.5,sloped, below] {$y_L^+=-\infty, ~w^+=-\infty$}; 
\draw (-8.5,2.5) node[below] {$i^L_{0}$};
\draw (-8.5,2.5)node [circle,fill,inner sep=1.5pt]{};
\draw (1.5,2.5) node[below] {$i^R_{0}$};
\draw (1.5,2.5)node [circle,fill,inner sep=1.5pt]{};
\draw (-3.5,2.5) node[below] {$q_{0}$};
\draw (-3.5,2.5) node [circle,fill,inner sep=1.5pt]{};
 \draw [very thick](1 mm, 10 pt) (-6,4.75)--(-1,.25) node[pos=.425,sloped, below] {$w^+=0, ~y_L^+=\infty\qquad \quad ~ y_R^+=-\infty$};
 \draw [very thick](1 mm, 10 pt) (-6,.25)--(-1,4.75)node[pos=.55,sloped, below] {$y_L^-=-\infty \qquad w^-=0,~y_R^-=\infty$};
 \draw[thick](1 mm, 10 pt) (-6,.25)--(-6,4.75) node[pos=.5,sloped, above] {$w^+w^-=-e^{\frac{4\pi \epsilon}{\beta}}$};
 \draw[thick](1 mm, 10 pt) (-1,.25)--(-1,4.75)node[pos=.5,sloped, below] {$w^+w^-=-e^{\frac{4\pi \epsilon}{\beta}}$};
\end{tikzpicture}
\caption{An eternal black hole in equilibrium with a finite temperature bath can be described using a plane with light cone coordinates $(w^+,w^-)$. The right/left Rindler wedge describes the right/left side of the black hole exterior coupled to the right/left bath having flat metric. The right/left Rindler wedge can be covered using the light cone coordinates $(y_{R/L}^+,y_{R/L}^-)$. The lines $w^+w^-=0$ are the future and the past horizons of the eternal black hole. The interface between the black hole and the bath satisfy the equation $w^+w^-=-e^{\frac{4\pi \epsilon}{\beta}}$. The red dashed hyperbola represents the singularity on which the dilaton profile vanishes.} \label{fig:BHAdS2bath}
\end{figure}

 Let us discuss the setup of an eternal black hole solution in JT gravity coupled to a non-gravitating bath. The bath and the black hole have the same inverse temperature $\beta$. The glued spacetime can be described as a region in a plane with light cone coordinates $(w^+,w^-)$ \cite{Almheiri:2019yqk}, see figure \ref{fig:BHAdS2bath}. The right Rindler wedge in the $w$-plane is the union of the right exterior of the black hole and the bath. Similarly, the union of the left side of the black hole and the bath forms the left Rindler wedge. The right Rindler wedge can be covered with the Minkowski coordinates $(y_R^+,y_R^-)$ of the right bath and the left Rindler wedge can be covered with the Minkowski coordinates $(y_L^+,y_L^-)$ of the left bath. They are related to the $(w^+,w^-)$ light cone coordinates as follows 
 \begin{equation}\label{BHbathwplane}
 w^{\pm}=\pm e^{\pm \frac{2\pi y_{R}^{\pm}}{\beta}}, \qquad w^{\pm}=\mp e^{\mp \frac{2\pi y_{L}^{\pm}}{\beta}}.
 \end{equation}
The locus of the black hole horizon is given by the equation $w^+w^-=0$. The singularity lies on the hyperbola on which the dilaton profile vanishes. The dilaton profile in this coordinates takes the following form 
\begin{equation}\label{dilaton}
\phi(w^+,w^-)=\phi_0+\frac{2\pi \phi_r}{\beta}\frac{1-w^+w^-}{1+w^+w^-},
\end{equation}
where $\phi_0$ gives rise to the extremal entropy and $\frac{\phi_r}{\epsilon}$ is the boundary value of the difference $\phi(w^+,w^-)-\phi_0$ at the boundary of the black hole spacetime given by the equation 
\begin{equation} 
w^+w^-=-e^{\frac{4\pi \epsilon}{\beta}}.
\end{equation}
 The real parameter $\epsilon$ specifies the location of the cut-out boundary. The metric in the AdS$_2$ region is 
\begin{equation} 
ds^2_{BH}=\frac{4dw^+dw^-}{(1+w^+w^-)^2},
\end{equation}
and the metric in the bath region is 
\begin{equation} 
ds^2_{B}=\frac{\beta^2}{4\pi^2\epsilon^2}\frac{dw^+dw^-}{w^+w^-}.
\end{equation}

\subsection{The appearance of an island and interior reconstruction}
\label{infoparadox}
 The classical geometry of an AdS$_2$ eternal black hole in equilibrium with a thermal bath is the same as that of an AdS$_2$ eternal black hole with reflecting boundary conditions at the boundary. However, quantum mechanically the two systems are significantly different. The constant exchange of the Hawking radiation and the radiation from the bath gives rise to an ever-growing entanglement between the black hole and the bath. This can be seen by computing the entanglement entropy of the bath. \par
 
 Assume that the matter conformal field theory in the glued spacetime has central charge $c$. Then the von Neumann entropy of the bath computed using the replica method \cite{Calabrese:2004eu} is given by
\begin{align}\label{bathentropy}
S_{\text{bath}}(u)=S_{\text{BH}}(u)&=\frac{c}{3}~\text{ln}\left(\frac{\beta}{\pi\epsilon}\text{cosh}\left(\frac{2\pi u}{\beta}\right)\right)\xrightarrow{u>>\beta} \frac{2\pi c }{3\beta} u,
\end{align}
 where $u=\frac{y^{+}-y^-}{2}$ is the time measured in the flat coordinates. The entanglement entropy of the bath/black hole increases linearly at late time. The unbounded growth makes the entanglement entropy bigger than the thermal entropy of the black hole at sufficiently large times. This paradoxical growth of the von Neumann entropy can be resolved by performing the entropy computation after choosing the saddle of the gravitational replica path integral having minimum free energy \cite{Almheiri:2019yqk, Almheiri:2019qdq}. At the initial times the trivial or the Hawking saddle has the minimum free energy. However, after the Page time, the time at which the von Neumann entropy of the black hole is equal to its Bekenstein-Hawking entropy, the saddle having the least free energy is a Euclidean wormhole geometry. The replica wormhole introduces a non-trivial entanglement wedge island, containing the black hole interior, for the bath. The appearance of an island for the bath after the Page time demands that the `reduced' half-sided translations defined using the operator algebras restricted to the AdS$_2$ region outside the horizon must ceases to translate a field in the gravitational region outside the horizon to the interior.

\subsection{The relative commutants and the reduced half-sided translations}
\label{seeisland}
\begin{figure}
\centering
\begin{tikzpicture}[scale=1.4, every node/.style={scale=.85}]
 \draw [red, very thick] plot [smooth] coordinates {(-6,4.75) (-3.5,3.5) (-1,4.75)} ;
 \draw [thick,](1 mm, 10 pt) (-3.5,7)--(1.5,2.5) ;
 \draw [thick,](1 mm, 10 pt) (-3.5,-2)--(1.5,2.5) ;
 \draw [red, very thick] plot[smooth]coordinates {(-6,0.25) (-3.5,1.5) (-1,0.25)} ;
 \draw [thick,](1 mm, 10 pt) (-3.5,7)--(-8.5,2.5) ;
 \draw [thick,](1 mm, 10 pt) (-3.5,-2)--(-8.5,2.5); 
 \fill[fill=black!50!green] (-2.15,3.5)--(-1.6,2.95)--(-1,3.5)--(-1.55,4.05);
 \fill[fill=blue!20] (-2,3.35)--(-1.6,2.95)--(-1,3.5)--(-1.4,3.9);
 \draw [blue, very thick,dashed] plot[smooth]coordinates {(-8.5,2.5)(-6,3.5) (-4.85,3.5) (-3.5,2.5) (-2.15,3.5) (-1,3.5)(1.5,2.5) } ;
\draw (-.75,3.6) node {$b_u^R$};
\draw (-1,3.5) node [circle,fill,inner sep=1.5pt]{};
\draw (-6,3.625) node[left] {$b_u^L$};
\draw (-6,3.5) node [circle,fill,inner sep=1.5pt]{};
\draw (-4.8,3.4) node[left] {$\Sigma_u$};
\draw (-8.5,2.5) node[below] {$i^L_{0}$};
\draw (-8.5,2.5)node [circle,fill,inner sep=1.5pt]{};
\draw (1.5,2.5) node[below] {$i^R_{0}$};
\draw (1.5,2.5)node [circle,fill,inner sep=1.5pt]{};
\draw (-1.8,3.35) node[below] {$q_{{N}_u}$};
\draw (-2,3.35)node [circle,fill,inner sep=.75pt]{};
\draw (-2.4,3.7) node[below] {$q_{{M}_u}$};
\draw (-3.5,2.5) node[below] {$q_{0}$};
\draw (-3.5,2.5) node [circle,fill,inner sep=1.5pt]{};
\draw (-2.15,3.5) node [circle,fill,inner sep=.75pt]{};
 \draw [very thick](1 mm, 10 pt) (-6,4.75)--(-1,.25) ;
 \draw [very thick](1 mm, 10 pt) (-6,.25)--(-1,4.75);
 \draw[thick](1 mm, 10 pt) (-6,.25)--(-6,4.75);
 \draw[thick](1 mm, 10 pt) (-1,.25)--(-1,4.75);
\end{tikzpicture}
\caption{ The half-sided translations are defined using the von Neumann algebra $\mathcal{M}_u$ restricted to $\Diamond_{M_u}$, the union of the green shaded and the blue shaded wedges in the right exterior of the eternal black hole and the subalgebra $\mathcal{N}_u$ restricted to $\Diamond_{N_u}$, the blue shaded wedge. The wedges $\Diamond_{M_u}$ and $\Diamond_{N_u}$ are the domain of dependence of two intervals $M_u$ and $N_u$ on the slice $\Sigma_u$} \label{fig:BHAdS2bathMN}
\end{figure}
 
 \begin{figure}
\centering
\begin{tikzpicture}[scale=1.4, every node/.style={scale=.85}]
 \fill[fill=yellow] (1.5,2.5)--(.825,1.9)--(-1,3.5)--(-.35,4.15);
 \fill[fill=yellow] (-5.7,0)--(-8.5,2.5)--(-4.8,5.8)--(-2,3.325);
 \draw [red, very thick] plot [smooth] coordinates {(-6,4.75) (-3.5,3.5) (-1,4.75)} ;
 \draw [thick,](1 mm, 10 pt) (-3.5,7)--(1.5,2.5);
 \draw [thick,](1 mm, 10 pt) (-3.5,-2)--(1.5,2.5);
 \draw [red, very thick] plot[smooth]coordinates {(-6,0.25) (-3.5,1.5) (-1,0.25)} ;
 \draw [thick,](1 mm, 10 pt) (-3.5,7)--(-8.5,2.5) ;
 \draw [thick,](1 mm, 10 pt) (-3.5,-2)--(-8.5,2.5); 
 \draw (-.75,3.6) node {$b_u^R$};
\draw (-1,3.5) node [circle,fill,inner sep=1.5pt]{};
\draw (-6,3.625) node[left] {$b_u^L$};
\draw (-6,3.5) node [circle,fill,inner sep=1.5pt]{};
\draw (-4.8,3.4) node[left] {$\Sigma_u$};
\draw (-8.5,2.5) node[below] {$i^L_{0}$};
\draw (-8.5,2.5)node [circle,fill,inner sep=1.5pt]{};
\draw (1.5,2.5) node[below] {$i^R_{0}$};
\draw (1.5,2.5)node [circle,fill,inner sep=1.5pt]{};
\draw (-1.8,3.35) node[below] {$q_{{N}_u}$};
\draw (-2,3.35)node [circle,fill,inner sep=.75pt]{};
\draw (-2.4,3.7) node[below] {$q_{{M}_u}$};
\draw (-2.15,3.5) node [circle,fill,inner sep=.75pt]{};
\draw (-3.5,2.5) node[below] {$q_{0}$};
\draw (-3.5,2.5) node [circle,fill,inner sep=1.5pt]{};
 \draw [blue, very thick,dashed] plot[smooth]coordinates {(-8.5,2.5)(-6,3.5) (-4.85,3.5) (-3.5,2.5) (-2.15,3.5) (-1,3.5)(1.5,2.5) } ;
 \draw [very thick](1 mm, 10 pt) (-6,4.75)--(-1,.25) ;
 \draw [very thick](1 mm, 10 pt) (-6,.25)--(-1,4.75);
 \draw[thick](1 mm, 10 pt) (-6,.25)--(-6,4.75) ;
 \draw[thick](1 mm, 10 pt) (-1,.25)--(-1,4.75);
\end{tikzpicture}
\caption{ The commutant $\mathcal{M}_u'$ of the von Neumann algebra $\mathcal{M}_u$ is given by the algebra of observables defined in the union of the three yellow shaded wedges.} \label{fig:BHAdS2bathMNC}
\end{figure}

Consider the equal time slice $\Sigma_u$ in $w$-plane, the glued spacetime. Suppose $\mathcal{M}_u$ be the algebra of operators restricted to the causal diamond $\Diamond_{M_u}$, the domain of dependence of the interval $(q_{M_u}, b^R_u)$ on $\Sigma_u$. It is the union of the green and blue shaded region in the right exterior, as shown in figure \ref{fig:BHAdS2bathMN}. The subalgebra $\mathcal{N}_u$ is the algebra of operators restricted to the causal diamond $\Diamond_{N_u}$, the blue shaded region in the right exterior which is a subregion of $\Diamond_{M_u}$. The causal diamond $\Diamond_{N_u}$ is the domain of dependence of the interval $N_u=(q_{N_u}, b_{u}^R)$ on $\tilde \Sigma_{u}$, a slice that is obtained by deforming $\Sigma_u$ without changing its points of intersection at the cut out boundaries. \par 

 Although, the algebras $\mathcal{M}_u$ and $\mathcal{N}_u$ are within the AdS$_2$ region, the respective commutants $\mathcal{M}'_u$ and $\mathcal{N}'_u$ are algebras that extend also to the non-gravitating bath. They are defined in the the domain of dependence of the interval $$M'_u=(i^L_0,q_0)\cup(q_0,q_{M_u})\cup (b_u^R,i^R_0)$$ and the domain of dependence of the interval $$N'_u=(i^L_0,q_0)\cup(q_0,q_{N_u})\cup (b_u^R,i^R_0)$$ respectively, see figure \ref{fig:BHAdS2bathMNC}. Here the point $q_0$ is the bifurcation point of the black hole. Hence, the translation of the operators in the gravitational region outside the black hole to the black hole interior by the action of $U(s,u)$ defined using the algebra $\mathcal{M}_u$ and subalgebra $\mathcal{N}_u$ can not be considered as a reconstruction of interior operators using the degrees of freedom within the gravitational region outside the black hole horizon. Therefore, the interior reconstruction via half-sided translation as described in \cite{Leutheusser:2021qhd,Leutheusser:2021frk} can not be used to demonstrate the island paradigm. Below, we will argue that the island paradigm can be demonstrated using a modified interior reconstruction method using the notion of the `reduced half-sided translations'. \par

 \begin{figure}
\centering
\begin{tikzpicture}[scale=1.4, every node/.style={scale=.85}]
 \fill[yellow] (-6,3.5) --(-4.05,1.55)--(-2,3.325)--(-4.05,5.25);
 \draw [red, very thick] plot [smooth] coordinates {(-6,4.75) (-3.5,3.5) (-1,4.75)} ;
 \draw [thick,](1 mm, 10 pt) (-3.5,7)--(1.5,2.5);
 \draw [thick,](1 mm, 10 pt) (-3.5,-2)--(1.5,2.5);
 \draw [red, very thick] plot[smooth]coordinates {(-6,0.25) (-3.5,1.5) (-1,0.25)} ;
 \draw [thick,](1 mm, 10 pt) (-3.5,7)--(-8.5,2.5) ;
 \draw [thick,](1 mm, 10 pt) (-3.5,-2)--(-8.5,2.5); 
 \draw (-.75,3.6) node {$b_u^R$};
\draw (-1,3.5) node [circle,fill,inner sep=1.5pt]{};
\draw (-6,3.625) node[left] {$b_u^L$};
\draw (-6,3.5) node [circle,fill,inner sep=1.5pt]{};
\draw (-4.8,3.4) node[left] {$\Sigma_u$};
\draw (-8.5,2.5) node[below] {$i^L_{0}$};
\draw (-8.5,2.5)node [circle,fill,inner sep=1.5pt]{};
\draw (1.5,2.5) node[below] {$i^R_{0}$};
\draw (1.5,2.5)node [circle,fill,inner sep=1.5pt]{};
\draw (-1.8,3.35) node[below] {$q_{{N}_u}$};
\draw (-2,3.35)node [circle,fill,inner sep=.75pt]{};
\draw (-2.4,3.7) node[below] {$q_{{M}_u}$};
\draw (-2.15,3.5) node [circle,fill,inner sep=.75pt]{};
\draw (-3.5,2.5) node[below] {$q_{0}$};
\draw (-3.5,2.5) node [circle,fill,inner sep=1.5pt]{};
 \draw [blue, very thick,dashed] plot[smooth]coordinates {(-8.5,2.5)(-6,3.5) (-4.85,3.5) (-3.5,2.5) (-2.15,3.5) (-1,3.5)(1.5,2.5) } ;
 \draw [very thick](1 mm, 10 pt) (-6,4.75)--(-1,.25) ;
 \draw [very thick](1 mm, 10 pt) (-6,.25)--(-1,4.75);
 \draw[thick](1 mm, 10 pt) (-6,.25)--(-6,4.75) ;
 \draw[thick](1 mm, 10 pt) (-1,.25)--(-1,4.75);
\end{tikzpicture}
\caption{ The relative commutant ${\mathcal{N}'}_{E_u}$ of the von Neumann algebra $\mathcal{N}_u$ within the AdS$_2$ region is given by algebra of observables defined in the yellow shaded wedge.} \label{fig:BHAdS2bathMNRC}
\end{figure}

The reduced half-sided translation $U^{E}(s,u)$, like $U(s,u)$, is defined with respect to the algebra $\mathcal{M}_u$ and the subalgebra $\mathcal{N}_u$. However, unlike $U(s,u)$, the reduced half-sided translation $U^E(s,u)$, instead of the commutants $\mathcal{M}_u'$ and $\mathcal{N}_u'$, uses the relative commutants \cite{Borchers:2000pv} $\mathcal{M}'_{E_u}$ and $\mathcal{N}'_{E_u}$ within the operator algebra $\mathcal{E}_u$ associated with the eternal black hole. The operator algebra associated with the black hole $\mathcal{E}_u$ is the operator algebra defined inside the domain of dependence of the interval $$E_u=(b_u^L,q_0)\cup(q_0,b_u^R)$$ on $\Sigma_u$. The relative commutant $\mathcal{M}'_{E_u}$ is the intersection of $\mathcal{M}'$ with $\mathcal{E}_u$ and the relative communtant $\mathcal{N}'_{E_u}$ is the intersection of $\mathcal{N}'$ with $\mathcal{E}_u$
\begin{equation}\label{relcomm}
 \mathcal{M}'_{E_u}=\mathcal{M}'\cap \mathcal{E}_u\qquad \mathcal{N}'_{E_u}=\mathcal{N}'\cap \mathcal{E}_u.
 \end{equation}
The algebra $\mathcal{M}'_{E_u}$ is the operator algebra defined in the domain of dependence of the interval $$M_{E_u}'=(b_u^L,q_0)\cup(q_0,q_{M_u}),$$ and the algebra $\mathcal{N}'_{E_u}$ is the union of all the observables in the domain of dependence of the interval $$N_{E_u}'=(b_u^L,q_0)\cup(q_0,q_{N_u}).$$ Therefore, the reduced half-sided translation $U^E(s,u)$ can be understood as the restriction of the half-sided translation $U(s,u)$ from the entire $w$-plane to the gravitational region. \par

The reduced half-sided translation can be explicitly expressed as 
\begin{equation}
\label{UsModopRed}
U^E(s,u)=\left({\Delta^E_{\mathcal{M}_u}}\right)^{-\text{i} t}\left({\Delta^E_{\mathcal{N}_u}}\right)^{\text{i} t}, \quad \quad s=e^{-2\pi t}-1 \quad \forall~ t\in \mathbb{R},
\end{equation}
where $\Delta^E_{\mathcal{M}_u}$ is the modular operator of the algebras $\mathcal{M}_u$ and the relative commutant $\mathcal{M}'_{E_u}$, and similarly for $\Delta^E_{\mathcal{N}_u}$. They are given by 
\begin{equation}
\Delta^E_{\mathcal{M}_u}={\rho^{-1}_{M'_{E_u}}}\rho_{{M_u}} \qquad \Delta^E_{\mathcal{N}_u}={\rho^{-1}_{N'_{E_u}}}\rho_{{N_u}}
\end{equation}
where $\rho_{M'_{E_u}}$ is the density matrix associated with the cut along the interval $M'_{E_u}$ and $\rho_{N'_{E_u}}$ is the density matrix associated with the cut along the interval $N'_{E_u}$. \par

\section{The reduced half-sided translation before Page time}\label{ARHSTBP}  

The action of reduced half-sided translation requires only knowing $\mathcal{K}_{\mathcal{M}_u}$ and $\mathcal{K}_{\mathcal{N}_u}$, the modular Hamiltonians associated with the operator algebras $\mathcal{M}_u$ and $\mathcal{N}_u$ respectively. They can be computed by following the method described in \cite{Cardy:2016fqc}. Assume that the algebras $\mathcal{M}_u$ and $\mathcal{N}_u$ are chosen such that $$w^+_{q_{M_u}}=w^+_{q_{N_u}}=w^-_{q_{M_u}}=0, \qquad w^-_{q_{N_u}}\neq 0.$$ For this choice of algebras the generator of the half-sided translations $U^{E}(s,u)$ is given by
\begin{equation}\label{ubhsT}
G^{E_u}=w^-_{q_{N_u}}\int_{-\infty}^{\infty} dz^-~T(z^-).
\end{equation}
It is the generator of translation along the $z^-$-direction, where $(z^+,z^-)$ are the coordinates whose entire range cover the causal diamond $\Diamond_{E_u}$, the domain of dependence of the interval $\left(b_u^L,b_u^R \right)$. The coordinates $(z^+,z^-)$ are related to the $w$-coordinates as follows
\begin{equation}\label{cuasalcord}
\left(z^+,z^-\right) = \begin{cases} \left(\frac{w^+_{b_u^R}\left(w^+-w^+_{q_{N_u}}\right)}{w^+_{b_u^R}-w^+}, \frac{w^-_{b_u^R}\left(w^--w^-_{q_{N_u}}\right)}{w^-_{b_u^R}-w^-}\right), & w^+<w^+_{q_{N_u}}, ~ w^-<w^-_{q_{N_u}}\\ \left(\frac{w^+_{b_u^L}\left(w^+-w^+_{q_{N_u}}\right)}{w^+_{b_u^L}-w^+}, \frac{w^-_{b_u^L}\left(w^--w^-_{q_{N_u}}\right)}{w^-_{b_u^L}-w^-}\right) &w^+>w^+_{q_{N_u}}, ~ w^->w^-_{q_{N_u}}\end{cases}. 
\end{equation}
Therefore, for appropriate values of $s$, the reduced half-sided translation $U^{E}(s,u)$ translate an operator outside the horizon to the black hole interior before the Page time.\par

\section{The reduced half-sided translation in quantum JT gravity}\label{RHSTJT}

 \begin{figure}
\centering
 \begin{tikzpicture}[scale=.7, every node/.style={scale=.7}]
 \fill[draw=black,fill=cyan!5] (-2,-2) rectangle (2,2);
 \fill[draw=black, fill=teal!30] (0,0) circle[radius=1.5];
 \draw[blue, very thick] (0,0.02) -- (0.3,0) to[out=10, in=150] (1.5,0);
 \draw[cyan,very thick] (0,-0.02) to[out=-160, in=-30] (-1.5,0);
 \draw[red, very thick] (0,-0.02) -- (0.3,0) to[out=-10, in=-150] (1.5,0);
 \draw[Orange, very thick] (0,0.02) to[out=160, in=30] (-1.5,0);
 \draw[Brown, very thick] (0.3,0) to[out=30, in=120] (1.5,0);
 \draw[Tan, very thick] (0.3,0) to[out=-150, in=-60] (-1.5,0);
 \draw[PineGreen, very thick] (0.3,0) to[out=-30, in=-120] (1.5,0);
 \draw[yellow, very thick] (0.3,0) to[out=150, in=60] (-1.5,0);
 \draw (0,0) node [circle,fill,inner sep=1pt]{};
 \draw (-0.2,.1) node [above]{$q_{{M}_u}$};
 \draw (0.3,0) node [circle,fill,inner sep=1.1pt]{};
 \draw (0.35,-0.1) node [below]{$q_{{N}_u}$};
 \draw (1.5,0) node [circle,fill,inner sep=1pt]{};
 \draw (1.5,0.2) node [right]{$b_u^R$};
 \draw (-1.5,0) node [circle,fill,inner sep=1pt]{};
 \draw (-2.15,0.2) node [right]{$b_u^L$};

 \fill[draw=black,fill=cyan!5] (-3,2) rectangle (-5,0);
 \fill[draw=black, fill=teal!30] (-4,1) circle[radius=0.75];
 \draw[blue, very thick] (-4,1.02) -- (-3.25,1);
 \draw (-4,1) node [circle,fill,inner sep=1pt]{};
 \draw (-4.2,1.1) node [above]{$q_{{M}_u}$};
 \draw (-3.25,1) node [circle,fill,inner sep=1pt]{};
 \draw (-2.5,1.2) node {$\rho_{{M}_u}$};

 \fill[draw=black,fill=cyan!5] (-5.5,2) rectangle (-7.5,0);
 \fill[draw=black, fill=teal!30] (-6.5,1) circle[radius=0.75];
 \draw[cyan,very thick] (-6.5,1.02) --(-7.25,1);
 \draw (-6.5,1) node [circle,fill,inner sep=1pt]{};
 \draw (-6.7,1.1) node [above]{$q_{{M}_u}$};
 \draw (-7.25,1) node [circle,fill,inner sep=1pt]{};
 \draw (-8,1.2) node {$\rho_{{M'_{E_u}}}$};

 \fill[draw=black,fill=cyan!5] (-3,4.5) rectangle (-5,2.5);
 \fill[draw=black, fill=teal!30] (-4,3.5) circle[radius=0.75];
 \draw[blue, very thick] (-4,3.52) -- (-3.25,3.5);
 \draw (-4,3.5) node [circle,fill,inner sep=1pt]{};
 \draw (-4.2,3.6) node [above]{$q_{{M}_u}$};
 \draw (-3.25,3.5) node [circle,fill,inner sep=1pt]{};
 \draw (-2.5,3.7) node {$\rho_{{M}_u}$};

 \fill[draw=black,fill=cyan!5] (-5.5,4.5) rectangle (-7.5,2.5);
 \draw[color=black, fill=teal!30](-6.5,3.5) circle[radius=0.75];
 \draw[cyan,very thick] (-6.5,3.52)-- (-7.25,3.5);
 \draw (-6.5,3.5) node [circle,fill,inner sep=1pt]{};
 \draw (-6.7,3.6) node [above]{$q_{{M}_u}$};
 \draw (-7.25,3.5) node [circle,fill,inner sep=1pt]{};
 \draw (-8,3.7) node {$\rho_{{M'_{E_u}}}$};

 \fill[draw=black,fill=cyan!5] (5.5,2) rectangle (7.5,0);
 \fill[draw=black, fill=teal!30] (6.5,1) circle[radius=0.75];
 \draw[red, very thick] (6.5,1.02) -- (7.25,1);
 \draw (6.5,1) node [circle,fill,inner sep=1pt]{};
 \draw (6.3,1.1) node [above]{$q_{{M}_u}$};
 \draw (7.25,1) node [circle,fill,inner sep=1pt]{};
 \draw (7.5,1.2) node [right]{$\rho_{{M}_u}$};

 \fill[draw=black,fill=cyan!5] (3,2) rectangle (5,0);
 \fill[draw=black, fill=teal!30] (4,1) circle[radius=0.75];
 \draw[Orange, very thick] (4,1.02) --(3.25,1);
 \draw (4,1) node [circle,fill,inner sep=1pt]{};
 \draw (3.8,1.1) node [above]{$q_{{M_u}}$};
 \draw (3.25,1) node [circle,fill,inner sep=1pt]{};
 \draw (2.5,1.2) node {$\rho_{{M'_{E_u}}}$};

 \fill[draw=black,fill=cyan!5] (5.5,4.5) rectangle (7.5,2.5);
 \draw[color=black, fill=teal!30](6.5,3.5) circle[radius=0.75];
 \draw[red, very thick] (6.5,3.52) -- (7.25,3.5);
 \draw (6.5,3.5) node [circle,fill,inner sep=1pt]{};
 \draw (6.3,3.6) node [above]{$q_{{M_u}}$};
 \draw (7.25,3.5) node [circle,fill,inner sep=1pt]{};
 \draw (7.5,3.7) node [right]{$\rho_{{M}_u}$};

 \fill[draw=black,fill=cyan!5] (3,4.5) rectangle (5,2.5);
 \fill[draw=black, fill=teal!30] (4,3.5) circle[radius=0.75];
 \draw[Orange, very thick] (4,3.52)--(3.25,3.5);
 \draw (4,3.5) node [circle,fill,inner sep=1pt]{};
 \draw (3.8,3.6) node [above]{$q_{{M}_u}$};
 \draw (3.25,3.5) node [circle,fill,inner sep=1pt]{};
 \draw (2.5,3.7) node {$\rho_{{M'_{E_u}}}$};

 \fill[draw=black,fill=cyan!5] (-3,7.5) rectangle (-5,5.5);
 \fill[draw=black, fill=teal!30] (-4,6.5) circle[radius=0.75];
 \draw[Brown, very thick] (-3.85,6.5) --(-3.25,6.5);
 \draw (-3.85,6.5) node [circle,fill,inner sep=1.1pt]{};
 \draw (-3.9,6.4) node [below]{$q_{{N}_u}$};
 \draw (-3.25,6.5) node [circle,fill,inner sep=1pt]{};
 \draw (-3,6.7) node [right]{$\rho_{{N}_u}$};

 \fill[draw=black,fill=cyan!5] (-5.5,7.5) rectangle (-7.5,5.5);
 \draw[color=black, fill=teal!30](-6.5,6.5) circle[radius=0.75];
 \draw[Tan, very thick] (-6.35,6.5)--(-7.25,6.5);
 \draw (-6.35,6.5) node [circle,fill,inner sep=1.1pt]{};
 \draw (-6.4,6.4) node [below]{$q_{{N}_u}$};
 \draw (-7.25,6.5) node [circle,fill,inner sep=1pt]{};
 \draw (-8,6.7) node {$\rho_{{N'_{E_u}}}$};

 \fill[draw=black,fill=cyan!5] (-3,10) rectangle (-5,8);
 \fill[draw=black, fill=teal!30] (-4,9) circle[radius=0.75];
 \draw[Brown, very thick] (-3.85,9) -- (-3.25,9);
 \draw (-3.85,9) node [circle,fill,inner sep=1.1pt]{};
 \draw (-3.9,8.9) node [below]{$q_{{N_u}}$};
 \draw (-3.25,9) node [circle,fill,inner sep=1pt]{};
 \draw (-3,9.2) node [right]{$\rho_{{N}_u}$};

 \fill[draw=black,fill=cyan!5] (-5.5,10) rectangle (-7.5,8);
 \draw[color=black, fill=teal!30](-6.5,9) circle[radius=0.75];
 \draw[Tan, very thick] (-6.35,9) --(-7.25,9);
 \draw (-6.35,9) node [circle,fill,inner sep=1.1pt]{};
 \draw (-6.4,8.9) node [below]{$q_{{N_u}}$};
 \draw (-7.25,9) node [circle,fill,inner sep=1pt]{};
 \draw (-8,8.9) node {$\rho_{{N'_{E_u}}}$};

 \fill[draw=black,fill=cyan!5] (5.5,7.5) rectangle (7.5,5.5);
 \draw[color=black, fill=teal!30](6.5,6.5) circle[radius=0.75];
 \draw[PineGreen, very thick] (6.65,6.5) -- (7.25,6.5);
 \draw (6.65,6.5) node [circle,fill,inner sep=1.1pt]{};
 \draw (6.7,6.4) node [below]{$q_{{N}_u}$};
 \draw (7.25,6.5) node [circle,fill,inner sep=1pt]{};
 \draw (7.5,6.7) node [right]{$\rho_{{N}_u}$};

 \fill[draw=black,fill=cyan!5] (3,7.5) rectangle (5,5.5);
 \fill[draw=black, fill=teal!30] (4,6.5) circle[radius=0.75];
 \draw[yellow, very thick] (4.15,6.5) --(3.25,6.5);
 \draw (4.15,6.5) node [circle,fill,inner sep=1.1pt]{};
 \draw (4.2,6.4) node [below]{$q_{{N}_u}$};
 \draw (3.25,6.5) node [circle,fill,inner sep=1pt]{};
 \draw (2.5,6.7) node {$\rho_{{N'_{E_u}}}$};

 \fill[draw=black,fill=cyan!5] (5.5,10) rectangle (7.5,8);
 \draw[color=black, fill=teal!30](6.5,9) circle[radius=0.75];
 \draw[PineGreen, very thick,middlearrow={>}] (6.65,9) -- (7.25,9);
 \draw (6.65,9) node [circle,fill,inner sep=1.1pt]{};
 \draw (6.7,8.9) node [below]{$q_{{N}_u}$};
 \draw (7.25,9) node [circle,fill,inner sep=1pt]{};
 \draw (7.5,9.2) node [right]{$\rho_{{N}_u}$};

 \fill[draw=black,fill=cyan!5] (3,10) rectangle (5,8);
 \fill[draw=black, fill=teal!30] (4,9) circle[radius=0.75];
 \draw[yellow, very thick] (4.15,9) -- (3.25,9);
 \draw (4.15,9) node [circle,fill,inner sep=1.1pt]{};
 \draw (4.2,8.9) node [below]{$q_{{N_u}}$};
 \draw (3.25,9) node [circle,fill,inner sep=1pt]{};
 \draw (2.5,8.9) node {$\rho_{{N'_{E_u}}}$};
 \end{tikzpicture}
\caption{The trivial saddle is obtained by cyclically gluing the $w$-planes along the cuts having same colours. The resulting Riemann surface viewed as a sheeted geometry over the $w$-plane has branch points at locations $b_u^R$ and $b_u^L$ on the $w$-plane. The figure describes the gluing for $n_1=n_2=2$.} \label{fig:USinpathintegral}
\end{figure}

 Consider the correlation function of the reduced half-sided translated operator $\mathcal{O}_s$ obtained by acting with $U^{E_u}(s)$ on $\mathcal{O}$, an element of the operator algebra $\mathcal{M}_u$
 $$\langle \Omega| {U^{E_u}(s)}^{\dagger}\mathcal{O}U^{E_u}(s) |\Omega\rangle$$ 
 where $|\Omega\rangle$ is the Hartle-Hawking state prepared by the Euclidean path integral. The nature of the action of $U^{E}(s,u)$ on $\mathcal{O}$ can be analysed by investigating this correlation function. The expression of $U^{E_u}(s)$ in terms of the density matrices enables us to identify the path integral description of this correlation function 
 \begin{equation}
\label{UsModopReddensityM}
U^{E_u}(s)=\left(\rho_{M'_{E_u}}^{-1}\rho_{{M}_u}\right)^{-\text{i} t}\left(\rho_{N'_{E_u}}^{-1}\rho_{{N}_u}\right)^{\text{i} t}, \quad \quad s=e^{-2\pi t}-1 \quad \forall~ t\in \mathbb{R}.
\end{equation} Substituting this gives that 
 \begin{align}\label{uEus}
 \langle \Omega| {U^{E_u}(s)}^{\dagger}\mathcal{O}U^{E_u}(s) |\Omega\rangle= \langle \Omega| \left(\rho_{N'_{E_u}}^{-1}\rho_{{N}_u}\right)^{-\text{i} t}\left(\rho_{M'_{E_u}}^{-1}\rho_{{M_u}}\right)^{\text{i} t}\mathcal{O} \left(\rho_{M'_{E_u}}^{-1}\rho_{{M_u}}\right)^{-\text{i} t}\left(\rho_{N'_{E_u}}^{-1}\rho_{{N_u}}\right)^{\text{i} t} |\Omega\rangle 
 \end{align} 
 This correlation function $$\langle \Omega| \rho_{{N}_u}^{n_1}\rho_{N'_{E_u}}^{n_2}\rho_{{M}_u}^{n_2}\rho_{M'_{E_u}}^{n_1}\mathcal{O}\rho_{M'_{E_u}}^{n_2}\rho_{{M_u}}^{n_1}\rho_{N'_{E_u}}^{n_1}\rho_{{N_u}}^{n_2} |\Omega\rangle$$ which can be given a replica like path integral description can reproduce the correlation function \eqref{uEus} if $n_1$ is analytically continued to $-\text{i}t$ and $n_2$ is analytically continued to $\text{i}t$. The path integral description of the Euclidean correlation function can be obtained by replacing each density matrix in the correlation function by its path integral prescription. \par
 
The density matrix $\rho_{A_u}$ associated with the interval ${A_u}$ is the matter conformal field theory euclidean path integral over a punctured $w$-plane with a cut from the puncture at the point $q_{A_u}$ to the AdS$_2$ cut-out boundary point $b_u^R$, for $A=M,N$. Similarly, the density matrix $\rho_{{B}_{E_u}'}$ is the matter conformal field theory euclidean path integral over the punctured $w$-plane with a cut from the puncture at the point $q_{B_u}$ to the AdS$_2$ cut-out boundary point $b_u^L$, for $B=M,N$. Using these prescriptions the correlation function \eqref{uEus} can be identified with the correlation function $$\langle \mathcal{O}\rangle_{\mathcal{G}^T_{n}}$$ of the matter CFT on the two dimensional surface $\mathcal{G}^T_{n}$. This surface is obtained by gluing $$n=4(n_1+n_2)+1$$ number of $w$-planes along the cuts on it as shown in figure \ref{fig:USinpathintegral}.\par 

Assume that the notion of reduced half-sided translation ${U}^{E_u}(s)$ can be extended to the full quantum theory of JT gravity coupled to CFT. Let $\mathcal{U}^{E_u}(s)$ be the reduced half-sided translation in the quantum JT gravity. Then path integral of CFT defined on $\mathcal{G}_{8n+1}^T$, for $n_1=-n_2$, can be understood as the semiclassical approximation of the correlation function:
\begin{equation}\label{necorrelQJT}
\lim_{G_N\to 0}\langle {\mathcal{U}^{E_u}(s,u)}^{\dagger} \mathcal{O}~\mathcal{U}^{E_u}(s,u) \rangle_{\mathbf{JT}},
 \end{equation}
where the subscript emphasises that the correlation function is computed in the quantum JT gravity coupled to the matter CFT. The path integral representation of this JT gravity correlation function can be expressed as follows 
\begin{equation} \label{necorrelp}
\langle {\mathcal{U}^{E_u}(s)}^{\dagger} \mathcal{O}~\mathcal{U}^{E_u}(s) \rangle_{\mathbf{JT}}= \lim_{n_2\to -n_1}\sum_{\substack{g_i, m_i\\ \sum_ig_i=g\quad\\ \sum_i m_i=n}}\int_{\mathcal{V}_{g,n}}\prod_j dl_jd\theta_j ~e^{-S_{JT}\left(\{l_j\theta_j\}\right)} \langle \mathcal{O} \rangle_{\mathcal{G}_{g,n}}
\end{equation}
where the surface $\mathcal{G}_{g,n}$ is obtained by replacing the gravitational region of $\mathcal{G}_{n}^{T}$ with a collection of hyperbolic Riemann surfaces $\left\{\mathcal{R}_{g_i,m_i}\right\}$ having total genus $g$ and $n$ boundaries with appropriate number of cuts on it. These cuts must be appropriately glued to each other to obtain the surface denoted as $\tilde{\mathcal{R}}_{g,n}$. Moreover the geometry near the boundary of the gravitational region must match with that of the gravitational region in $\mathcal{G}^{T}_n$. The space $\mathcal{V}_{g,n}$ is the subspace of the moduli space of the hyperbolic Riemann surface $\tilde{\mathcal{R}}_{g,n}$. The special feature of the subspace $\mathcal{V}_{g,n}$ is that the path integral satisfy all the constraints, if there is any, imposed by the requirement that the operator $\mathcal{U}^{E_u}(s)$ form a continuous group at least in the $G_N\to 0$ limit. The JT gravity action $S_{JT}\left(\{l^i_j,\theta^i_j\}\right)$ is a function of the Fenchel-Nielsen coordinates $\left(\cdots l_i,\theta_{j}\cdots \right)$ of the space $\mathcal{V}_{g,n}$. \par

The dominant saddle in the gravitational path integral \eqref{necorrelp} can be determined by minimising the free energy of the geometries contributing to it. Since the island paradigm is expected to be not dependent crucially on the details of the matter CFT, we assume that minimising the universal terms in the free energy is sufficient to determine the saddle. Before the Page time, the dominant saddle is the the geometry $\mathcal{G}^{n,T}$. It is obtained by sewing the $n$ number of $w$-planes along various cuts as shown in figure \ref{fig:USinpathintegral}. The resulting surface has eight conical singularities at points $b_u^R$ and $b_u^L$ which is located at the boundary of the gravitational region in the $w$-planes. The universal contribution to the free energy of CFT defined on a Riemann surface with a conical singularity having excess angle $\theta$ is given by \cite{Cardy:1988tk} 
\begin{equation}\label{wedgefenergy}
 F\approx c\Theta\left(\theta\right)\text{ln}~L,
\end{equation}
where $L$ can be understood as the radius of the largest flat neighbourhood around the conical singularity which does not include any other conical singularities and $$\Theta\left(\theta\right)=\frac{(\theta+2\pi)}{24 \pi}\left( 1-\left(\frac{2\pi}{(\theta+2\pi)} \right)^2\right).$$ Therefore, the universal term in the free energy of the trivial saddle is given by the sum of the contributions from all conical singularities. It is given by
\begin{equation}\label{trivialsaddleFE}
 F_{T}\approx \sum_{i=1}^2\frac{(n_i+1)c}{3}\left( 1-\frac{1}{(n_i+1)^2} \right)\text{ln}\left(\frac{\beta}{\pi\epsilon}\text{cosh}\left(\frac{2\pi u}{\beta}\right)\right).
\end{equation}
Here $L$ is the coordinate distance between the conical singularities (the distance between the points $b_u^L$ and $b_u^R$). After the Lorentzian continuation $L$ takes the form
\begin{equation}
\label{charlength}
L=\sqrt{(w^+_{b_u^R}-w^+_{b_u^L})(w^-_{b_u^R}-w^-_{b_u^L})}.
\end{equation}
 The Lorentzian coordinates of the points $b_L$ and $b_R$ at boundary time $u$ are given by
\begin{align}\label{BuRLcoordinates}
(w^+_{b_L},w^-_{b_L})&=\left(-e^{-\frac{2\pi (u-\epsilon)}{\beta}},e^{\frac{2\pi (u+\epsilon)}{\beta}}\right),\nonumber\\
(w^+_{b_R},w^-_{b_R})&=\left(e^{\frac{2\pi (u-\epsilon)}{\beta}},-e^{\frac{-2\pi (u+\epsilon)}{\beta}}\right)
\end{align}
Therefore, the time dependence of approximate free energy of the trivial saddle that describe the action of the half-sided translation $U^{BH}(s)$ for large $u$ is given by 
\begin{equation}\label{fenergycloseT}
{F}_{T}(u)\approx \sum_{i=1}^2\frac{2\pi c n_i(n_i+2)}{3\beta(n_i+1)}u.
\end{equation}
At late times, the free energy of the trivial saddle \eqref{fenergycloseT} grows linearly with respect to the boundary time. Therefore, we expect that at late times there can be other saddles having free energy less than that of the trivial saddle. 

\section{The reduced half-sided translations after the Page time}\label{RHSTAPT}

In this section, we shall construct a wormhole saddle that satisfies the JT gravity constraint and contributes to the path integral \eqref{necorrelp} with free energy less than that of the trivial saddle after the Page time. A non-trivial saddle can have bridges connecting various sheets that are being glued along the cuts. Since the matter CFT is not coupled to the JT gravity dilaton, the constraint demands that the wormhole must have curvature $R=-2$, i.e. the wormhole must be a hyperbolic Riemann surface. Our aim is to construct such a wormhole by gluing hyperbolic quadrilaterals. We will then compute the free energy of the wormhole geometries, and minimise it to identify the non-trivial saddle. The wormhole saddle like the trivial saddle can be described as a sheeted geometry with each sheet being a $w$-plane with cuts in the gravitational region. We will argue that this can be interpreted as a correlation function of an operator acted by a modified reduced half-sided translation. Finally, we will study the action of this modified reduced half-sided translation. 
 
\subsection{Construction of a wormhole geometry}
 \begin{figure}
\centering
 \begin{tikzpicture}[scale=.7]
 \draw[teal!30,fill=teal!20]plot [smooth] coordinates {(-3.25,12)(-2,9)(0,8)(2,9)(3.25,12)(3.4,12.5)(4,12.75)(4.5,12.55)(4.75,12.2)(4.7,11.65)(4.7,11.65)(3.5,9.5)(3,8.5)(3,8.5)(4,9.5)(5.8,12.3)(5.8,12.3)(5.9,12.5)(6.5,12.75)(7,12.55)(7.25,12.2)(7.1,11.55)(7.1,11.55)(6.5,11)(4.85,8.5)(5,8)(6,8.7)(8.45,10.55)(8.45,10.55)(9,10.75)(9.6,10.5)(9.75,9.8)(9.5,9.45)(9.4,9.35)(9.3,9.3)(9.3,9.3)(5.2,7.25)(4.5,6.7)(6.2,7.1)(8.7,8.2)(8.7,8.2)(9,8.25)(9.6,8)(9.75,7.3)(9.5,6.95)(9.1,6.75)(8.9,6.75)(8.9,6.75)(6,6.2)(5,5.5)(6.4,5)(8.7,5.2)(8.7,5.2)(9,5.25)(9.6,5)(9.6,5)(9.75,4.3)(9.5,3.95)(9,3.75)(8.7,3.8)(8.7,3.8)(5.5,4.5)(4.5,4.5)(9.1,2.75)(9.1,2.75)(9.6,2.5)(9.75,1.8)(9.5,1.45)(9,1.25)(8.7,1.3)(8.7,1.3)(4,4)(2.75,4.25)(3,3.85)(7.1,.5)(7.1,0.5)(7.25,-.2)(7,-.55)(6.5,-.75)(5.95,-.5)(5.95,-.5)(4.05,1.85)(3.75,2.2)(3.25,2.55)(3,2.3)(4.65,0.35)(4.65,0.35)(4.75,-.2)(4.5,-.55)(4,-.75)(3.4,-.5)(3.25,0)(1.5,2.9)(1,3.4)(.7,3)(1.5,0) (1.2,.5)(1.5,0)(-1.5,0)(-.7,3)(-1,3.4)(-1.5,2.9)(-3.25,0)(-3.25,0)(-3.4,-.5)(-4,-.75)(-4.5,-.55)(-4.75,-.2)(-4.65,0.35)(-4.65,0.35)(-3,2.3)(-3.25,2.55)(-3.75,2.2)(-4.05,1.85)(-5.95,-.5)(-5.95,-.5)(-6.5,-.75)(-7,-.55)(-7.25,-.2)(-7.1,0.5)(-7.1,.5)(-4,3)(-3,3.85)(-2.75,4.25)(-4,4)(-8.7,1.3)(-9,1.25)(-9.5,1.45)(-9.75,1.8)(-9.6,2.5)(-9.1,2.75)(-9.1,2.75)(-4.5,4.5)(-5.5,4.5)(-6.5,4.25)(-8.7,3.8)(-8.7,3.8)(-9,3.75)(-9.5,3.95)(-9.75,4.3)(-9.6,5)(-9,5.25)(-8.7,5.2)(-8.7,5.2)(-6.4,5)(-5,5.5)(-6,6.2)(-8.9,6.75)(-8.9,6.75)(-9.1,6.75)(-9.5,6.95)(-9.75,7.3)(-9.6,8)(-9,8.25)(-8.7,8.2)(-8.7,8.2)(-6.2,7.1)(-4.5,6.7)(-5.2,7.25)(-9.3,9.3)(-9.3,9.3)(-9.4,9.35)(-9.5,9.45)(-9.75,9.8)(-9.6,10.5)(-9,10.75)(-8.45,10.55)(-8.45,10.55)(-6,8.7)(-5,8)(-4.85,8.5)(-6.5,11)(-7.1,11.55)(-7.1,11.55)(-7.25,12.2)(-7,12.55)(-6.5,12.75)(-5.9,12.5)(-5.8,12.3)(-5.8,12.3)(-4,9.5)(-3,8.5)(-3.5,9.5)(-4.7,11.65)(-4.7,11.65)(-4.75,12.2)(-4.5,12.55)(-4,12.75)(-3.4,12.5)(-3.25,12)} ; 
 
 \fill[draw=black,fill=cyan!5] (-2,-2) rectangle (2,2);
 \fill[draw=black, fill=white!20] (0,0) circle[radius=1.5];
 \draw[Brown, thick] plot [smooth] coordinates{(-3.7,6.6)(-2.5,6.3)(-.6,5)(-0.4,2)(-1.5,0)};
 \draw[cyan, thick ]plot [smooth] coordinates{ (-3.5,5)(-1.8,5.1)(-0.7,4)(-0.75,2)(-1.5,0)};
 \draw[blue, thick, densely dotted, ]plot [smooth] coordinates{(-2.3,3.15)(-1.9,4) (-.95,4.4)(0.15,2)(1.5,0)};
 \draw[Orange, thick] plot [smooth] coordinates{(3.5,5)(2.7,5.2)(1,4.5)(0,1)(-1.5,0)};
 \draw[PineGreen, thick] plot [smooth] coordinates{(3.7,6.6)(.4,6)(0.2,3)(-.1,1.6)(-1.5,0)};
 \draw[Tan, very thick,densely dotted] plot [smooth] coordinates{ (-2.5,7.9)(-2,7.1)(-.2,5.55)(.2,3)(0.5,2)(1.5,0)};;
 \draw[red, thick,densely dotted] plot [smooth] coordinates{ (2.3,3.2)(1.75,4)(.75,4.3)(0.55,4)(.6,3)(0.85,2)(1.5,0)};
 \draw[violet, thick,densely dotted] plot [smooth] coordinates{(2.5,8)(.2,5.1)(0,3)(.2,2)(1.5,0)};
 \draw (1.5,0) node [circle,fill,inner sep=1pt]{};
 \draw (.55,2.3) node [right]{$\mathcal{O}$};
 \draw (1.4,0.2) node [right]{$b_u^R$};
 \draw (-1.5,0) node [circle,fill,inner sep=1pt]{};
 \draw (-2.5,0.2) node [right]{$b_u^L$};
 \draw (.5,2) node [circle,fill,inner sep=1pt]{};
 \fill[draw=black,fill=cyan!5] (-3,1) rectangle (-5,-1);
 \fill[draw=black, fill=white!20] (-4,0) circle[radius=0.75];
 \draw[blue, thick, densely dotted] plot [smooth] coordinates{(-2.3,3.25)(-2.65,1.5)(-3.25,0)};
 \draw (-4.5,-1.5) node [right]{$\rho_{N}$};
 \draw (3.5,-1.5) node [right]{$\rho_{N}$};
 \draw (-7,13.5) node [right]{$\rho_{M}$};
 \draw (6,13.5) node [right]{$\rho_{M}$};

 \fill[draw=black,fill=cyan!5] (-5.5,1) rectangle (-7.5,-1);
 \fill[draw=black, fill=white!20] (-6.5,0) circle[radius=0.75];
 \draw[blue, thick, densely dotted] plot [smooth] coordinates{(-2.3,3.25) (-4.25,2)(-5.75,0)};
 \draw (-11.5,2) node [right]{$\rho_{N'_{E_u}}$};
 \draw (10,2) node [right]{$\rho_{N'_{E_u}}$};
 \draw (-11.5,7.5) node [right]{$\rho_{M'_{E_u}}$};
 \draw (10,7.5) node [right]{$\rho_{M'_{E_u}}$};

 \fill[draw=black,fill=cyan!5] (-8,3) rectangle (-10,1);
 \fill[draw=black, fill=white!20] (-9,2) circle[radius=0.75];
 \draw[cyan, thick] plot [smooth] coordinates{(-3.5,4.8)(-4,4.2)(-6,3.3)(-7.5,2.6) (-9.75,2)};
 \draw (-7.5,-1.5) node [right]{$\rho_{N}$};
 \draw (5.75,-1.5) node [right]{$\rho_{N}$};
 \draw (2.75,13.5) node [right]{$\rho_{M}$};
 \draw (-4.5,13.5) node [right]{$\rho_{M}$};

 \fill[draw=black,fill=cyan!5] (-8,5.5) rectangle (-10,3.5);
 \draw[color=black, fill=white!20](-9,4.5) circle[radius=0.75];
 \draw[cyan, thick] plot [smooth] coordinates{(-3.5,4.8)(-4,5.4)(-7,4.7)(-8.5,4.6)(-9.75,4.5)};
 \draw (-11.5,4.5) node [right]{$\rho_{N'_{E_u}}$};
 \draw (10.1,4.5) node [right]{$\rho_{N'_{E_u}}$};
 \draw (10.1,10) node [right]{$\rho_{M'_{E_u}}$};
 \draw (-11.5,10) node [right]{$\rho_{M'_{E_u}}$};

 \fill[draw=black,fill=cyan!5] (5.5,1) rectangle (7.5,-1);
 \fill[draw=black, fill=white!20] (6.5,0) circle[radius=0.75];
 \draw[red, thick,densely dotted] plot [smooth] coordinates{(2.2,3.2)(2.95,3)(6,.75)(7.25,0)};

 \fill[draw=black,fill=cyan!5] (3,1) rectangle (5,-1);
 \fill[draw=black, fill=white!20] (4,0) circle[radius=0.75];
 \draw[red, thick,densely dotted] plot [smooth] coordinates{(2.2,3.2)(2.5,2)(4.75,0)};

 \fill[draw=black,fill=cyan!5] (8,5.5) rectangle (10,3.5);
 \draw[color=black, fill=white!20](9,4.5) circle[radius=0.75];
 \draw[orange, thick] plot [smooth] coordinates{(3.5,5)(4,5.1)(7.5,4.6)(8.25,4.5)};

 \fill[draw=black,fill=cyan!5] (8,3) rectangle (10,1);
 \fill[draw=black, fill=white!20] (9,2) circle[radius=0.75];
 \draw[orange, thick] plot [smooth] coordinates{(3.5,5)(6,3.25) (8.25,2)};

 \fill[draw=black,fill=cyan!5] (-8,11) rectangle (-10,9);
 \fill[draw=black, fill=white!20] (-9,10) circle[radius=0.75];
 \draw[Brown, thick]plot [smooth] coordinates{(-3.7,6.6)(-5,7.8)(-7.75,9.4)(-9.75,10)};

 \fill[draw=black,fill=cyan!5] (-8,8.5) rectangle (-10,6.5);
 \draw[color=black, fill=white!20](-9,7.5) circle[radius=0.75];
 \draw[Brown, thick] plot [smooth] coordinates{(-3.7,6.6)(-4.5,6.5)(-7,7) (-9.75,7.5)};

 \fill[draw=black,fill=cyan!5] (-3,13) rectangle (-5,11);
 \fill[draw=black, fill=white!20] (-4,12) circle[radius=0.75];
 \draw[Tan, thick,densely dotted] plot [smooth] coordinates{(-2.5,7.8)(-3.35,10.5)(-3.25,12)};

 \fill[draw=black,fill=cyan!5] (-5.5,13) rectangle (-7.5,11);
 \draw[color=black, fill=white!20](-6.5,12) circle[radius=0.75];
 \draw[Tan, thick, densely dotted] plot [smooth] coordinates{(-2.5,7.8)(-4,8.5)(-5.75,12)};
 
 \fill[draw=black,fill=cyan!5] (8,11) rectangle (10,9);
 \draw[color=black, fill=white!20](9,10) circle[radius=0.75];
 \draw[PineGreen, thick]plot [smooth] coordinates{(3.7,6.6)(5,7.8)(7.75,9.4)(8.25,10)};

 \fill[draw=black,fill=cyan!5] (8,8.5) rectangle (10,6.5);
 \fill[draw=black, fill=white!20] (9,7.5) circle[radius=0.75];
 \draw[PineGreen, thick] plot [smooth] coordinates{(3.7,6.6)(4.5,6.5)(7,7) (8.25,7.5)};

 \fill[draw=black,fill=cyan!5] (5.5,13) rectangle (7.5,11);
 \draw[color=black, fill=white!20](6.5,12) circle[radius=0.75];
 \draw[violet, thick,densely dotted] plot [smooth] coordinates{(2.5,8)(4.5,9.5) (6.5,11.5) (7.25,12)};

 \fill[draw=black,fill=cyan!5] (3,13) rectangle (5,11);
 \fill[draw=black, fill=white!20] (4,12) circle[radius=0.75];
 \draw[violet, thick, densely dotted] plot [smooth] coordinates{(2.5,8)(3.35,10.5)(4.75,12)};
 \draw [black,dashed, very thick ]plot [smooth] coordinates {(3.5,6.7)(3.2,7)(3.9,7.5)(5,8)} ;
 \draw [black,dashed, very thick ]plot [smooth] coordinates {(3.5,6.7)(3.2,6.5)(3.9,6.3)(5.5,6)} ;
 \draw [black,very thick]plot [smooth] coordinates {(3.7,6.6)(3.2,6.8)(3.9,7.3)(5,8)} ;
 \draw [black,very thick]plot [smooth] coordinates {(3.7,6.6)(3.2,6.4)(3.9,6.2)(5.5,6)} ;
 \draw [black,dashed, very thick ]plot [smooth] coordinates {(-3.5,6.7)(-3.2,7)(-3.9,7.5)(-5,8)} ;
 \draw [black,dashed, very thick ]plot [smooth] coordinates {(-3.5,6.7)(-3.2,6.5)(-3.9,6.3)(-5.5,6)} ;
 \draw [black,very thick]plot [smooth] coordinates {(-3.7,6.6)(-3.2,6.8)(-3.9,7.3)(-5,8)} ;
 \draw [black,very thick]plot [smooth] coordinates {(-3.7,6.6)(-3.2,6.4)(-3.9,6.2)(-5.5,6)} ;
 \draw [black,dashed, very thick ]plot [smooth] coordinates {(-2.5,8)(-1.7,7.5)(-1.5,7.6)(-1.4,8.5)} ;
 \draw [black,very thick]plot [smooth] coordinates {(-2.5,7.8)(-1.7,7.4)(-1.4,7.5)(-1.4,8.5)} ;
 \draw [black,dashed, very thick ]plot [smooth] coordinates {(2.5,8)(1.7,7.5)(1.5,7.6)(1.4,8.5)} ;
 \draw [black,very thick]plot [smooth] coordinates {(2.5,7.8)(1.7,7.4)(1.4,7.5)(1.4,8.5)} ;
 \draw [dashed,thick]plot [smooth] coordinates {(2.5,8)(2.7,8.1)(3,8.5)} ;
 \draw [thick]plot [smooth] coordinates {(2.5,7.8)(3.1,8.3)(3,8.5)} ;
 \draw [dashed,thick]plot [smooth] coordinates {(-2.5,8)(-2.7,8.1)(-3,8.5)} ;
 
 \draw [thick]plot [smooth] coordinates {(-2.5,7.8)(-3.1,8.3)(-3,8.5)} ;
 
 \draw [dashed,very thick]plot [smooth] coordinates {(2.5,8)(3,7.65)(4.8,8.5)} ;
 \draw [very thick]plot [smooth] coordinates {(2.5,7.8)(3,7.5)(4.8,8.5)} ;
 
 \draw [dashed,very thick]plot [smooth] coordinates {(-2.5,8)(-3,7.65)(-4.8,8.5)} ;
 \draw [very thick]plot [smooth] coordinates {(-2.5,7.8)(-3,7.5)(-4.8,8.5)} ;
 
 \draw [dashed,thick]plot [smooth] coordinates {(3.5,6.7)(4.2,6.9)(4.45,6.7)} ;
 \draw [thick]plot [smooth] coordinates {(3.7,6.6)(4.2,6.7)(4.45,6.7)} ;
 \draw [dashed,thick]plot [smooth] coordinates {(-3.5,6.7)(-4.2,6.9)(-4.45,6.7)} ;
 \draw [thick]plot [smooth] coordinates {(-3.7,6.6)(-4.2,6.7)(-4.45,6.7)} ;
 \draw [black,dashed, very thick]plot [smooth] coordinates {(2.3,3.2)(1.3,3.8)(1,3.5)(.8,3.5)(.85,3.4)} ;
 \draw [black,very thick]plot [smooth] coordinates {(2.5,3.3)(1.5,4)(.95,4)(.85,3.4)} ;
 \draw [black,dashed, very thick]plot [smooth] coordinates {(-2.3,3.2)(-1.3,3.8)(-1,3.5)(-.8,3.5)(-.85,3.4)} ;
 \draw [black,very thick]plot [smooth] coordinates {(-2.5,3.3)(-1.5,4)(-.95,4)(-.85,3.4)} ;
 
 \draw [black,dashed, very thick]plot [smooth] coordinates {(3.5,4.8)(2.9,5.7)(3.6,5.9)(4.4,5.8)(5,5.5)} ;
 \draw [black,very thick]plot [smooth] coordinates {(3.5,5)(3.1,5.6)(4.2,5.8)(5,5.5)} ;
 \draw [black,dashed, very thick]plot [smooth] coordinates {(-3.5,4.8)(-2.9,5.7)(-3.6,5.9)(-4.4,5.8)(-5,5.5)} ;
 \draw [black,very thick]plot [smooth] coordinates {(-3.5,5)(-3.1,5.6)(-4.2,5.8)(-5,5.5)} ;
 
 \draw [dashed, very thick]plot [smooth] coordinates {(3.25,4.3)(2.7,4.5)(2.3,5)(3.5,4.8)} ;
 \draw [dashed, very thick]plot [smooth] coordinates {(-3.25,4.3)(-2.7,4.5)(-2.3,5)(-3.5,4.8)} ;
 \draw [dashed, very thick]plot [smooth] coordinates {(2.3,3.2)(2,4.3)(2.75,4.1)} ;
 \draw [dashed, very thick]plot [smooth] coordinates {(-2.3,3.2)(-2,4.3)(-2.75,4.1)} ;
 \draw [very thick]plot [smooth] coordinates {(3.25,4.3)(2.5,4.5)(2.5,4.8)(3.5,5)} ;
 \draw [very thick]plot [smooth] coordinates {(2.5,3.3)(2.2,4)(2.75,4.1)} ;
 \draw [very thick]plot [smooth] coordinates {(-3.25,4.3)(-2.5,4.5)(-2.5,4.8)(-3.5,5)} ;
 \draw [very thick]plot [smooth] coordinates {(-2.5,3.3)(-2.2,4)(-2.75,4.1)} ;
 
 \draw [dashed, thick]plot [smooth] coordinates {(3.5,4.8)(4,4.6)(4.5,4.6)} ;
 \draw [thick]plot [smooth] coordinates {(3.5,5)(4,4.9)(4.5,4.6)} ;
 \draw [dashed, thick]plot [smooth] coordinates {(-3.5,4.8)(-4,4.6)(-4.5,4.6)} ;
 \draw [thick]plot [smooth] coordinates {(-3.5,5)(-4,4.9)(-4.5,4.6)} ;
 
 \draw [black]plot [smooth] coordinates {(3.25,0)(1.5,2.9)(1,3.4)(.7,3)(1.5,0)} ;
 \draw [black]plot [smooth] coordinates {(-1.5,0)(-.7,3)(-1,3.4)(-1.5,2.9)(-3.25,0)} ;
 \draw [black]plot [smooth] coordinates {(-4.65,0.35)(-3,2.3)(-3.25,2.55)(-3.75,2.2)(-4.05,1.85)(-5.95,-.5)};
 \draw [black]plot [smooth] coordinates {(-7.1,.5)(-4,3)(-3,3.85)(-2.75,4.25)(-4,4)(-8.7,1.3)} ;
 \draw [black]plot [smooth] coordinates {(-9.1,2.75)(-4.5,4.5)(-5.5,4.5)(-6.5,4.25)(-8.7,3.8)};
 \draw [black]plot [smooth] coordinates {(-8.7,5.2)(-6.4,5)(-5,5.5)(-6,6.2)(-8.9,6.75)};
 \draw [black]plot [smooth] coordinates {(-8.7,8.2)(-6.2,7.1)(-4.5,6.7)(-5.2,7.25)(-9.3,9.3)};
 \draw [black]plot [smooth] coordinates {(-8.45,10.55)(-6,8.7)(-5,8)(-4.85,8.5)(-6.5,11)(-7.1,11.55)};
 \draw [black]plot [smooth] coordinates {(-5.8,12.3)(-4,9.5)(-3,8.5)(-3.5,9.5)(-4.7,11.65)};
 \draw [black]plot [smooth] coordinates {(3.25,12)(2,9)(0,8)(-2,9)(-3.25,12)} ;
 \draw [black]plot [smooth] coordinates {(4.65,0.35)(3,2.3)(3.25,2.55)(3.75,2.2)(4.05,1.85)(5.95,-.5)};
 \draw [black]plot [smooth] coordinates {(7.1,.5)(4,3)(3,3.85)(2.75,4.25)(4,4)(8.7,1.3)} ;
 \draw [black]plot [smooth] coordinates {(9.1,2.75)(4.5,4.5)(5.5,4.5)(6.5,4.25)(8.7,3.8)};
 \draw [black]plot [smooth] coordinates {(8.7,5.2)(6.4,5)(5,5.5)(6,6.2)(8.9,6.75)};
 \draw [black]plot [smooth] coordinates {(8.7,8.2)(6.2,7.1)(4.5,6.7)(5.2,7.25)(9.3,9.3)};
 \draw [black]plot [smooth] coordinates {(8.45,10.55)(6,8.7)(5,8)(4.85,8.5)(6.5,11)(7.1,11.55)};
 \draw [black]plot [smooth] coordinates {(5.8,12.3)(4,9.5)(3,8.5)(3.5,9.5)(4.7,11.65)};
 
 \end{tikzpicture}
\caption{A wormhole geometry with with $n=4n_1+4n_2+1$ number of boundaries with copies of bath glued to them. The gravitational region of this geometry has the topology of a sphere with $n$ number of boundaries and additional gluing along the cuts indicated by the coloured dotted and thick lines on it. None of the coloured lines on the surface cross each other. All the cuts having the same colours are being cyclically glued. The gravitational region of the wormhole geometry has smooth hyperbolic metric. The coloured dotted and the thick coloured lines indicate cuts. Here, the value of $n_1$ and $n_2$ is chosen to be 2.} \label{fig:USinpathintegralwormholeswmap}
\end{figure}

 It is an elementary fact in hyperbolic geometry that the gluing of two hyperbolic Riemann surfaces produces another hyperbolic Riemann surface if and only if they are glued along a closed geodesic or along an open geodesic that connects two cusps or conical singularities in the hyperbolic Riemann surfaces \cite{imayoshi2012introduction,pan2017finite}. This implies that a cut in the wormhole geometry must be an open geodesic on the surface. Using these facts, we shall attempt to identify a wormhole geometry having a smooth hyperbolic metric on it. \par

\begin{figure}
\centering
 \begin{tikzpicture}[scale=1]
 
 \fill[draw=black,fill=cyan!5] (-2,-2) rectangle (2,2);
 \fill[draw=black, fill=white!20] (0,0) circle[radius=1.5];
 \draw[Brown, thick] plot [smooth] coordinates{(-3.7,6.6)(-2.5,6.3)(-.6,5)(-0.4,2)(-1.5,0)};
 \draw[cyan, thick ]plot [smooth] coordinates{ (-3.5,5)(-1.8,5.1)(-0.7,4)(-0.75,2)(-1.5,0)};
 \draw[blue, thick, densely dotted, ]plot [smooth] coordinates{(-2.3,3.15)(-1.9,4) (-.95,4.4)(0.15,2)(1.5,0)};
 \draw[Orange, thick] plot [smooth] coordinates{(3.5,5)(2.7,5.2)(1,4.5)(0,1)(-1.5,0)};
 \draw[PineGreen, thick] plot [smooth] coordinates{(3.7,6.6)(.4,6)(0.2,3)(-.1,1.6)(-1.5,0)};
 \draw[Tan, very thick,densely dotted] plot [smooth] coordinates{ (-2.5,7.9)(-2,7.1)(-.2,5.55)(.2,3)(0.5,2)(1.5,0)};;
 \draw[red, thick,densely dotted] plot [smooth] coordinates{ (2.3,3.2)(1.75,4)(.75,4.3)(0.55,4)(.6,3)(0.85,2)(1.5,0)};
 \draw[violet, thick,densely dotted] plot [smooth] coordinates{(2.5,8)(.2,5.1)(0,3)(.2,2)(1.5,0)};

 \draw (1.5,0) node [circle,fill,inner sep=1pt]{};
 \draw (.55,2.3) node [right]{$\mathcal{O}$};
 \draw (1.4,0.2) node [right]{$b_u^R$};
 \draw (-1.5,0) node [circle,fill,inner sep=1pt]{};
 \draw (-2.1,0.2) node [right]{$b_u^L$};
 \draw (.5,2) node [circle,fill,inner sep=1pt]{};

 \draw [black,dashed, thick ]plot [smooth] coordinates {(3.5,6.7)(3.2,7)(3.9,7.5)(5,8)} ;
 \draw [black,dashed, thick ]plot [smooth] coordinates {(3.5,6.7)(3.2,6.5)(3.9,6.3)(5.5,6)} ;
 \draw [black, thick ]plot [smooth] coordinates {(5,8)(4.75,8.5)} ;
 \draw [black, thick ]plot [smooth] coordinates {(-5,8)(-4.75,8.5)} ;
 \draw [black, thick ]plot [smooth] coordinates {(5.5,6)(5,5.5)} ;
 \draw [black, thick ]plot [smooth] coordinates {(-5.5,6)(-5,5.5)} ;
 \draw [black, thick ]plot [smooth] coordinates {(3.25,4.3)(2.75,4.1)} ;
 \draw [black, thick ]plot [smooth] coordinates {(-3.25,4.3)(-2.75,4.1)} ;
 \draw [black, thick]plot [smooth] coordinates {(3.7,6.6)(3.2,6.8)(3.9,7.3)(5,8)} ;
 \draw [black, thick]plot [smooth] coordinates {(3.7,6.6)(3.2,6.4)(3.9,6.2)(5.5,6)} ;
 \draw (3.7,6.6) node [circle,fill,inner sep=1pt]{};
 \draw (4,6.4) node []{$\tilde q_6$};
 \draw (3.5,6.7) node [circle,fill,inner sep=1pt]{};
 \draw (3.8,6.9) node []{$ q_6$};
 \draw (-3.7,6.6) node [circle,fill,inner sep=1pt]{};
 \draw (-4,6.4) node []{$\tilde q_3$};
 \draw (-3.5,6.7) node [circle,fill,inner sep=1pt]{};
 \draw (-3.8,6.9) node []{$ q_3$};
 \draw [black,dashed, thick ]plot [smooth] coordinates {(-3.5,6.7)(-3.2,7)(-3.9,7.5)(-5,8)} ;
 \draw [black,dashed, thick ]plot [smooth] coordinates {(-3.5,6.7)(-3.2,6.5)(-3.9,6.3)(-5.5,6)} ;
 \draw [black, thick]plot [smooth] coordinates {(-3.7,6.6)(-3.2,6.8)(-3.9,7.3)(-5,8)} ;
 \draw [black, thick]plot [smooth] coordinates {(-3.7,6.6)(-3.2,6.4)(-3.9,6.2)(-5.5,6)} ;
 \draw [black,dashed, very thick ]plot [smooth] coordinates {(-2.5,8)(-1.7,7.5)(-1.5,7.6)(-1.4,8.5)} ;
 \draw [black, thick]plot [smooth] coordinates {(-2.5,7.8)(-1.7,7.4)(-1.4,7.5)(-1.4,8.5)} ;
 \draw [black,dashed, thick ]plot [smooth] coordinates {(2.5,8)(1.7,7.5)(1.5,7.6)(1.4,8.5)} ;
 \draw [black, thick]plot [smooth] coordinates {(2.5,7.8)(1.7,7.4)(1.4,7.5)(1.4,8.5)} ;
 
 \draw [dashed, thick]plot [smooth] coordinates {(2.5,8)(3,7.65)(4.8,8.5)} ;
 \draw [ thick]plot [smooth] coordinates {(2.5,7.8)(3,7.5)(4.8,8.5)} ;
 
 \draw [dashed, thick]plot [smooth] coordinates {(-2.5,8)(-3,7.65)(-4.8,8.5)} ;
 \draw [ thick]plot [smooth] coordinates {(-2.5,7.8)(-3,7.5)(-4.8,8.5)} ;
 
 \draw [black,dashed, thick]plot [smooth] coordinates {(2.3,3.2)(1.3,3.8)(1,3.5)(.8,3.5)(.85,3.4)} ;
 \draw [black, thick]plot [smooth] coordinates {(2.5,3.3)(1.5,4)(.95,4)(.85,3.4)} ;
 \draw [black,dashed, thick]plot [smooth] coordinates {(-2.3,3.2)(-1.3,3.8)(-1,3.5)(-.8,3.5)(-.85,3.4)} ;
 \draw [black, thick]plot [smooth] coordinates {(-2.5,3.3)(-1.5,4)(-.95,4)(-.85,3.4)} ;
 
 \draw [black,dashed, thick]plot [smooth] coordinates {(3.5,4.8)(2.9,5.7)(3.6,5.9)(4.4,5.8)(5,5.5)} ;
 \draw [black, thick]plot [smooth] coordinates {(3.5,5)(3.1,5.6)(4.2,5.8)(5,5.5)} ;
 \draw [black,dashed, very thick]plot [smooth] coordinates {(-3.5,4.8)(-2.9,5.7)(-3.6,5.9)(-4.4,5.8)(-5,5.5)} ;
 \draw [black, thick]plot [smooth] coordinates {(-3.5,5)(-3.1,5.6)(-4.2,5.8)(-5,5.5)} ;
 \draw (3.5,5) node [circle,fill,inner sep=1pt]{};
 \draw (3.8,5.1) node []{$\tilde q_7$};
 \draw (3.5,4.8) node [circle,fill,inner sep=1pt]{};
 \draw (3.7,4.5) node []{$ q_7$};
 \draw (-3.5,5) node [circle,fill,inner sep=1pt]{};
 \draw (-3.8,5.1) node []{$\tilde q_2$};
 \draw (-3.5,4.8) node [circle,fill,inner sep=1pt]{};
 \draw (-3.7,4.5) node []{$ q_2$};
 
 \draw (2.3,3.2) node [circle,fill,inner sep=1pt]{};
 \draw (2.3,3) node []{$\tilde q_8$};
 \draw (2.5,3.3) node [circle,fill,inner sep=1pt]{};
 \draw (2.7,3.5) node []{$q_8$};
 \draw (-2.3,3.2) node [circle,fill,inner sep=1pt]{};
 \draw (-2.3,2.9) node []{$\tilde q_1$};
 \draw (-2.5,3.3) node [circle,fill,inner sep=1pt]{};
 \draw (-2.7,3.6) node []{$q_1$};
 
 \draw (-2.5,7.8) node [circle,fill,inner sep=1pt]{};
 \draw (-2.9,7.8) node []{$q_4$};
 \draw (-2.5,8) node [circle,fill,inner sep=1pt]{};
 \draw (-2.5,8.2) node []{$\tilde q_4$};
 \draw (2.5,7.8) node [circle,fill,inner sep=1pt]{};
 \draw (2.9,7.7) node []{$ q_5$};
 \draw (2.5,8) node [circle,fill,inner sep=1pt]{};
 \draw (2.9,8.1) node []{$ \tilde q_5$};
 
 \draw [dashed, thick]plot [smooth] coordinates {(3.25,4.3)(2.7,4.5)(2.3,5)(3.5,4.8)} ;
 \draw [dashed, thick]plot [smooth] coordinates {(-3.25,4.3)(-2.7,4.5)(-2.3,5)(-3.5,4.8)} ;
 \draw [dashed, thick]plot [smooth] coordinates {(2.3,3.2)(2,4.3)(2.75,4.1)} ;
 \draw [dashed, thick]plot [smooth] coordinates {(-2.3,3.2)(-2,4.3)(-2.75,4.1)} ;
 \draw [ thick]plot [smooth] coordinates {(3.25,4.3)(2.5,4.5)(2.5,4.8)(3.5,5)} ;
 \draw [ thick]plot [smooth] coordinates {(2.5,3.3)(2.2,4)(2.75,4.1)} ;
 \draw [ thick]plot [smooth] coordinates {(-3.25,4.3)(-2.5,4.5)(-2.5,4.8)(-3.5,5)} ;
 \draw [ thick]plot [smooth] coordinates {(-2.5,3.3)(-2.2,4)(-2.75,4.1)} ;
 
 \draw [black,thick]plot [smooth] coordinates {(1,3.4)(.7,3)(1.5,0)} ;
 \draw [black,thick]plot [smooth] coordinates {(-1.5,0)(-.7,3)(-1,3.4)} ;

 \draw [black,thick]plot [smooth] coordinates {(1.5,8.5)(0,8)(-1.5,8.5)} ;

 \draw (2.4,7.5) node [below]{$2\zeta_2$};
 \draw[->,very thick]plot [smooth] coordinates{(2.15,7.6) (2.3,7.4) (2.5,7.4)(2.65,7.6)};
 \draw (2.3,8.3) node [above]{$2\pi-2\gamma$};
 \draw[<-,very thick]plot [smooth] coordinates{(2.15,7.9) (2.3,8.3) (2.5,8.3)(2.65,7.9)};

 \end{tikzpicture}
\caption{ The `tree' is a hyperbolic disc with eight cuts. The interior angle at points $q_i, ~i=1,2,3,4$ is $2\zeta_1$ and the interior angle at points $q_i, ~i=5,6,7,8$ is $2\zeta_2$. Interior angle at points $\tilde q_i$ is $2\gamma$. Scissoring the surface along the coloured lines produces hyperbolic polygons.} \label{fig:USinpathintegralwormholeswmap13}
\end{figure}

\begin{figure}
\centering
\begin{tikzpicture}[scale=.7, every node/.style={scale=0.8}]
 \draw[teal!30,fill=teal!20]plot [smooth] coordinates {(-10,6.1) (-10,1)(-10,0)(-10.5,0)(-12.5,0)(-13,0)(-13,0.5) (-13,6.1)(-12.5,5)(-11.5,4.5)(-10.5,5.1)(-10,6.1) } ; 
 \draw[teal!30,fill=teal!20]plot [smooth] coordinates {(-5,6.1) (-5,1)(-5,0)(-5.5,0)(-7.5,0)(-8,0)(-8,0.5) (-8,6.1)(-7.5,5)(-6.5,4.5)(-5.5,5.1)(-5,6.1) } ; 
 \draw[teal!30,fill=teal!20]plot [smooth] coordinates {(-1.5,0)(-.75,3.8) (-.75,3.8) (-.5,4)(-.35,4.5) (-.3,6.1)(.15,4.5)(.5,4)(.75,3.8)(.75,3.8)(1.5,0)(0,1.5) (-1.5,0)} ; 
 \draw[teal!30,fill=teal!20]plot [smooth] coordinates {(-.75,3.8)(-.75,4)(-.5,4.5)(0.3,6.1)(.6,4.5)(.75,4)(.75,3.8)(-.75,3.8)} ;
 \draw[ thick,middlearrow={>}]plot [smooth] coordinates{(-10,6.1) (-10,0)};
 \draw[ middlearrow={<}, thick]plot [smooth] coordinates{(-13,6.1) (-13,0)};
 \draw[ thick]plot [smooth] coordinates{(-13,0) (-10,0)};
 \draw[ thick,middlearrow={<}]plot [smooth] coordinates{(-8,6.1) (-8,0)};
 \draw[ thick]plot [smooth] coordinates{(-13,6.1) (-12.5,5)(-11.5,4.5)(-10.5,5.1)(-10,6.1)};
 \draw[ thick]plot [smooth] coordinates{(-8,6.1) (-7.5,5)(-6.5,4.5)(-5.5,5.1)(-5,6.1)};
 \draw[ thick]plot [smooth] coordinates{(-8,0) (-5,0)};
 \draw[ middlearrow={>}, thick]plot [smooth] coordinates{(-5,6.1) (-5,0)};
 \draw[ thick,->]plot [smooth] coordinates{(-4.5,3) (-2,3)};
 \fill[draw=black, thick, fill=white!20] (0,0) circle[radius=1.5];
 \draw (-0.3,6.1) node [circle,fill,inner sep=1pt]{};
 \draw (0.3,6.1) node [circle,fill,inner sep=1pt]{};
 \draw (1.5,0) node [circle,fill,inner sep=1pt]{};
 \draw (1.5,0.2) node [right]{$B$};
 \draw (-1.5,0) node [circle,fill,inner sep=1pt]{};
 \draw (-2.4,0.2) node [right]{$A$};
 \draw (-0.3,6.1) node [left]{$C$};
 \draw (0.3,6.1) node [right]{$D$};
 \draw (-5,0.2) node [right]{$B'$};
 \draw (-5,6.2) node [right]{$C'$};
 \draw (-8,0.2) node [left]{$A'$};
 \draw (-8,6.2) node [left]{$D'$};
 \draw (-10,0.2) node [right]{$B''$};
 \draw (-10,6.2) node [right]{$C''$};
 \draw (-13,0.2) node [left]{$A''$};
 \draw (-13,6.2) node [left]{$D''$};
 \draw [black, densely dotted, very thick]plot [smooth] coordinates {(-.3,6.1)(.15,4.5)(.5,4)(.75,3.8)} ;
 \draw [black, thick]plot [smooth] coordinates {(0.3,6.1)(.6,4.5)(.75,4)(.75,3.8)} ;
 \draw [black, densely dotted, very thick]plot [smooth] coordinates {(-.3,6.1)(-.35,4.5)(-.5,4)(-.75,3.8)} ;
 \draw [black, thick]plot [smooth] coordinates {(0.3,6.1)(-.5,4.5)(-.75,4)(-.75,3.8)} ;
 \draw [black,thick]plot [smooth] coordinates {(.75,3.8)(1.5,0)} ;
 \draw [black,thick]plot [smooth] coordinates {(-1.5,0)(-.75,3.8)} ;
\end{tikzpicture}
\caption{ A hyperbolic disks having two conical singularities at $D$ having interior angle $2\alpha$ and at $C$ having interior angle $2\beta$ can be obtained by identifying the edges $A'C'$and $A''C''$ and the edges $B'D'$ and $B''D''$ of a pair of hyperbolic quadrilaterals. The vertices $C'$ and $C''$ of the quadrilaterals have equal angle $\beta$ and the vertices $D'$ and $D''$ have equal angle $\alpha$. } \label{fig:USinpathintegralwormholeswmap12}
\end{figure}

 Consider a wormhole geometry $\mathcal{G}_{n,0}$ described in figure \ref{fig:USinpathintegralwormholeswmap}. The gravitational region of this geometry, denoted as $\tilde{\mathcal{R}}_{0,n}$, is obtained by cyclically gluing the cuts indicated by the same coloured curves on the Riemann sphere with $n$ boundaries. Each boundary of the wormhole is attached to a copy of the Euclidean bath. Riemann surface $\tilde{\mathcal{R}}_{0,n}$ with hyperbolic metric on it can be constructed by identifying the boundaries of a set of hyperbolic polygons without introducing any conical singularities. In order to demonstrate this, let us scissor $\tilde{\mathcal{R}}_{0,n}$ as shown in figure \ref{fig:USinpathintegralwormholeswmap} along the curves shown as black thick and dotted curves. The scissoring gives $n$ number of component surfaces. Out of which, $n-1=4n_1+4n_2$ surfaces are `crowns', disks with a cut . They can be divided into eight groups as follows.
 \begin{itemize}
 \item \underline{Group 1}: $n_1$ number of disks with a cut connecting a bulk point denoted as $\tilde{p}_1$ and the boundary point $b_u^R$. 
 \item \underline{Group 2}: $n_1$ number of disks with a cut connecting a bulk point denoted as $\tilde{p}_2$ and the boundary point $b_u^L$. 
 \item \underline{Group 3}: $n_1$ number of disks with a cut connecting a bulk point denoted as $\tilde{p}_3$ and the boundary point $b_u^R$. 
 \item \underline{Group 4}: $n_1$ number of disks with a cut connecting a bulk point denoted as $\tilde{p}_4$ and the boundary point $b_u^L$.
 \item \underline{Group 5}: $n_2$ number of disks with a cut connecting a bulk point denoted as $\tilde{p}_5$ and the boundary point $b_u^R$. 
 \item \underline{Group 6}: $n_2$ number of disks with a cut connecting a bulk point denoted as $\tilde{p}_6$ and the boundary point $b_u^L$. 
 \item \underline{Group 7}: $n_2$ number of disks with a cut connecting a bulk point denoted as $\tilde{p}_7$ and the boundary point $b_u^L$
 \item \underline{Group 8}: $n_2$ number of disks with a cut connecting a bulk point denoted as $\tilde{p}_8$ and the boundary point $b_u^L$. 
 \end{itemize}
 
 The points $p_1,p_2,p_3$ and $p_4$ are assumed to have deficit angle of $2\pi-\alpha_1$ and the points $p_5,p_6,p_7$ and $p_8$ are assumed to have deficit angle of $2\pi-\alpha_2$, see figure \ref{fig:USinpathintegralwormholeswmap12}. Moreover, point $\tilde p_i$ on the $i^{\text{th}}$ cut has corners with interior angle $\beta$. Such a crown with hyperbolic metric on it can be constructed by gluing a pair of hyperbolic quadrilaterals whose edges are geodesics as shown in figure \ref{fig:USinpathintegralwormholeswmap12}. The remaining component is a `tree', disk with eight cuts, see figure \ref{fig:USinpathintegralwormholeswmap13}. The end points of the cuts $q_i, ~i=1,2,3,4$ have interior angle  $2\zeta_1$ and the interior angle at points $q_i, ~i=5,6,7,8$ is $2\zeta_2$. Moreover, two points on these cuts have corners with interior angle $\gamma$. Such a tree with hyperbolic metric on it also can be constructed by gluing hyperbolic quadrilaterals along the boundaries. This validates our claim that $\tilde{\mathcal{R}}_{0,n}$ with hyperbolic metric on it can be constructed by identifying the boundaries of a set of hyperbolic quadrilaterals. \par

Being able to construct the wormhole geometry by gluing the edges of hyperbolic quadrilaterals itself does not guarantee that the metric on it is a smooth hyperbolic metric. For this, we must also make sure that the gluing does not produce conical singularity at the locations on the resulting surface where the vertices of the hyperbolic quadrilaterals meet. Hence we must demand that the cycles which can be shrunk to the points where the vertices of the hyperbolic quadrilaterals meet has no excess or deficit angles. This condition constraints the angles of each of the hyperbolic quadrilaterals. In order to avoid the presence of conical singularities at any point on the surface, we must set 
\begin{align}\label{csingremoval}
2n_i\alpha_i+2\zeta_i=\gamma+\beta=2\beta=2\pi \qquad i=1,2.
\end{align} 
 
\subsection{The wormhole saddle}
 As described in the previous subsection, a wormhole geometry $\mathcal{G}_{n,0}$ can be obtained by gluing $n$ number of $w$-planes along the cuts on them. Therefore, wormhole saddle can be found by identifying the locations and the deficit angles at the end points of these cuts that are in the bulk of the gravitational region. For this we will minimise the free energy of the JT gravity coupled to CFT path integral on the wormhole geometry. Since the ability to reconstruct the black hole interior is not expected to be crucially dependent on the details of the matter CFT, we assume that the change of saddle after the Page time is dictated by the universal terms in the free energy of the matter CFT. The contribution to the free energy from a conical singularity of the geometry where the CFT is defined is an example for such universal terms. \par

\begin{figure}
\centering
 \begin{tikzpicture}[scale=1, every node/.style={scale=1}]
 \fill[draw=black,fill=cyan!5] (-2,-2) rectangle (2,2);
 \fill[draw=black, fill=teal!30] (0,0) circle[radius=1.5];
 \draw[blue, very thick] (0.3,0) to[out=50, in=130] (1.5,0);
 \draw[cyan,very thick] (-.3,0) to[out=-150, in=-50] (-1.5,0);
 \draw[red, very thick] (0.3,0) to[out=-50, in=-130] (1.5,0);
 \draw[Orange, very thick] (-0.3,0) to[out=130, in=50] (-1.5,0);
 \draw[Brown, very thick] (0.5,0) to[out=30, in=170] (1.5,0);
 \draw[Tan, very thick] (-0.5,0) to[out=-170, in=-30] (-1.5,0);
 \draw[PineGreen, very thick] (0.5,0) to[out=-30, in=-170] (1.5,0);
 \draw[yellow, very thick] (-0.5,0) to[out=170, in=30] (-1.5,0);
 \draw (-0.3,0) node [circle,fill,inner sep=1pt]{};
 \draw (-0.5,0) node [circle,fill,inner sep=1pt]{};
 \draw (0.5,.1) node [above]{$q_4$};
 \draw (-0.5,.1) node [above]{$q_3$};
 \draw (-0.75,.1) node [below]{$q_2$};
 \draw (0.3,0) node [circle,fill,inner sep=1.1pt]{};
 \draw (0.5,0) node [circle,fill,inner sep=1.1pt]{};
 \draw (0.75,-0.1) node [below]{$q_1$};
 \draw (1.5,0) node [circle,fill,inner sep=1pt]{};
 \draw (1.5,0.2) node [right]{$b_u^R$};
 \draw (-1.5,0) node [circle,fill,inner sep=1pt]{};
 \draw (-2.15,0.2) node [right]{$b_u^L$};

 \fill[draw=black,fill=cyan!5] (-3,2) rectangle (-5,0);
 \fill[draw=black, fill=teal!30] (-4,1) circle[radius=0.75];
 \draw[blue, very thick] (-3.8,1.02) -- (-3.25,1);
 \draw (-3.8,1) node [circle,fill,inner sep=1pt]{};
 \draw (-3.8,1.1) node [above]{$p_4$};
 \draw (-3.25,1) node [circle,fill,inner sep=1pt]{};

 \fill[draw=black,fill=cyan!5] (-5.5,2) rectangle (-7.5,0);
 \fill[draw=black, fill=teal!30] (-6.5,1) circle[radius=0.75];
 \draw[cyan,very thick] (-6.7,1.02) --(-7.25,1);
 \draw (-6.7,1) node [circle,fill,inner sep=1pt]{};
 \draw (-6.5,1.1) node [above]{$p_3$};
 \draw (-7.25,1) node [circle,fill,inner sep=1pt]{};

 \fill[draw=black,fill=cyan!5] (-3,4.5) rectangle (-5,2.5);
 \fill[draw=black, fill=teal!30] (-4,3.5) circle[radius=0.75];
 \draw[blue, very thick] (-3.8,3.52) -- (-3.25,3.5);
 \draw (-3.8,3.5) node [circle,fill,inner sep=1pt]{};
 \draw (-3.8,3.6) node [above]{$ p_4$};
 \draw (-3.25,3.5) node [circle,fill,inner sep=1pt]{};

 \fill[draw=black,fill=cyan!5] (-5.5,4.5) rectangle (-7.5,2.5);
 \draw[color=black, fill=teal!30](-6.5,3.5) circle[radius=0.75];
 \draw[cyan,very thick] (-6.7,3.52)-- (-7.25,3.5);
 \draw (-6.7,3.5) node [circle,fill,inner sep=1pt]{};
 \draw (-6.5,3.6) node [above]{$p_3$};
 \draw (-7.25,3.5) node [circle,fill,inner sep=1pt]{};

 \fill[draw=black,fill=cyan!5] (5.5,2) rectangle (7.5,0);
 \fill[draw=black, fill=teal!30] (6.5,1) circle[radius=0.75];
 \draw[red, very thick] (6.7,1.02) -- (7.25,1);
 \draw (6.7,1) node [circle,fill,inner sep=1pt]{};
 \draw (6.7,1.1) node [above]{$p_4$};
 \draw (7.25,1) node [circle,fill,inner sep=1pt]{};

 \fill[draw=black,fill=cyan!5] (3,2) rectangle (5,0);
 \fill[draw=black, fill=teal!30] (4,1) circle[radius=0.75];
 \draw[Orange, very thick] (3.8,1.02) --(3.25,1);
 \draw (3.8,1) node [circle,fill,inner sep=1pt]{};
 \draw (4,1.1) node [above]{$p_3$};
 \draw (3.25,1) node [circle,fill,inner sep=1pt]{};

 \fill[draw=black,fill=cyan!5] (5.5,4.5) rectangle (7.5,2.5);
 \draw[color=black, fill=teal!30](6.5,3.5) circle[radius=0.75];
 \draw[red, very thick] (6.7,3.52) -- (7.25,3.5);
 \draw (6.7,3.5) node [circle,fill,inner sep=1pt]{};
 \draw (6.5,3.6) node [above]{$p_4$};
 \draw (7.25,3.5) node [circle,fill,inner sep=1pt]{};

 \fill[draw=black,fill=cyan!5] (3,4.5) rectangle (5,2.5);
 \fill[draw=black, fill=teal!30] (4,3.5) circle[radius=0.75];
 \draw[Orange, very thick] (3.8,3.52)--(3.25,3.5);
 \draw (3.8,3.5) node [circle,fill,inner sep=1pt]{};
 \draw (4,3.6) node [above]{$p_3$};
 \draw (3.25,3.5) node [circle,fill,inner sep=1pt]{};

 \fill[draw=black,fill=cyan!5] (-3,7.5) rectangle (-5,5.5);
 \fill[draw=black, fill=teal!30] (-4,6.5) circle[radius=0.75];
 \draw[Brown, very thick] (-3.6,6.5) --(-3.25,6.5);
 \draw (-3.6,6.5) node [circle,fill,inner sep=1.1pt]{};
 \draw (-3.7,6.4) node [below]{$p_1$};
 \draw (-3.25,6.5) node [circle,fill,inner sep=1pt]{};

 \fill[draw=black,fill=cyan!5] (-5.5,7.5) rectangle (-7.5,5.5);
 \draw[color=black, fill=teal!30](-6.5,6.5) circle[radius=0.75];
 \draw[Tan, very thick] (-6.9,6.5)--(-7.25,6.5);
 \draw (-6.9,6.5) node [circle,fill,inner sep=1.1pt]{};
 \draw (-6.6,6.4) node [below]{$p_2$};
 \draw (-7.25,6.5) node [circle,fill,inner sep=1pt]{};

 \fill[draw=black,fill=cyan!5] (-3,10) rectangle (-5,8);
 \fill[draw=black, fill=teal!30] (-4,9) circle[radius=0.75];
 \draw[Brown, very thick] (-3.6,9) -- (-3.25,9);
 \draw (-3.6,9) node [circle,fill,inner sep=1.1pt]{};
 \draw (-3.7,8.9) node [below]{$p_1$};
 \draw (-3.25,9) node [circle,fill,inner sep=1pt]{};

 \fill[draw=black,fill=cyan!5] (-5.5,10) rectangle (-7.5,8);
 \draw[color=black, fill=teal!30](-6.5,9) circle[radius=0.75];
 \draw[Tan, very thick] (-6.9,9) --(-7.25,9);
 \draw (-6.9,9) node [circle,fill,inner sep=1.1pt]{};
 \draw (-6.6,8.9) node [below]{$p_2$};
 \draw (-7.25,9) node [circle,fill,inner sep=1pt]{};

 \fill[draw=black,fill=cyan!5] (5.5,7.5) rectangle (7.5,5.5);
 \draw[color=black, fill=teal!30](6.5,6.5) circle[radius=0.75];
 \draw[PineGreen, very thick] (6.9,6.5) -- (7.25,6.5);
 \draw (6.9,6.5) node [circle,fill,inner sep=1.1pt]{};
 \draw (6.8,6.4) node [below]{$p_1$};
 \draw (7.25,6.5) node [circle,fill,inner sep=1pt]{};

 \fill[draw=black,fill=cyan!5] (3,7.5) rectangle (5,5.5);
 \fill[draw=black, fill=teal!30] (4,6.5) circle[radius=0.75];
 \draw[yellow, very thick] (3.6,6.5) --(3.25,6.5);
 \draw (3.6,6.5) node [circle,fill,inner sep=1.1pt]{};
 \draw (3.9,6.4) node [below]{$p_2$};
 \draw (3.25,6.5) node [circle,fill,inner sep=1pt]{};

 \fill[draw=black,fill=cyan!5] (5.5,10) rectangle (7.5,8);
 \draw[color=black, fill=teal!30](6.5,9) circle[radius=0.75];
 \draw[PineGreen, very thick] (6.9,9) -- (7.25,9);
 \draw (6.9,9) node [circle,fill,inner sep=1.1pt]{};
 \draw (6.8,8.9) node [below]{$p_1$};
 \draw (7.25,9) node [circle,fill,inner sep=1pt]{};

 \fill[draw=black,fill=cyan!5] (3,10) rectangle (5,8);
 \fill[draw=black, fill=teal!30] (4,9) circle[radius=0.75];
 \draw[yellow, very thick] (3.6,9) -- (3.25,9);
 \draw (3.6,9) node [circle,fill,inner sep=1.1pt]{};
 \draw (3.9,8.9) node [below]{$p_2$};
 \draw (3.25,9) node [circle,fill,inner sep=1pt]{};
 \end{tikzpicture}
\caption{The special class of wormhole geometries ${\mathcal{G}}_{n,0}$ is obtained by cyclically gluing the $w$-planes along the cuts having same colours. The figure describes the gluing for $n_1=n_2=2$.} \label{fig:USinpathintegralW}
\end{figure}

 The universal terms in the free energy can be computed by scissoring the wormhole geometry $\mathcal{G}_{n,0}$ along the black curves shown in figure \ref{fig:USinpathintegralwormholeswmap}. The resulting surface $\tilde{\mathcal{G}}_{n,0}$ can be constructed by gluing the $n-1$ number of $w$-planes having a cut, and a $w$-plane having eight cuts along a part of these cuts. In order to specify the gluing more precisely, let us consider the following nine groups of $w$ planes with cuts.
 
 \begin{itemize}
 \item \underline{Group 1}: $n_1$ number of $w$-planes with a cut connecting two bulk points of the gravitational region denoted as $p_1$ and $\tilde{p}_1$ and the boundary point $b_u^R$ of the gravitational region. 
 \item \underline{Group 2}: $n_1$ number of $w$-planes with a cut connecting two bulk points of the gravitational region denoted as $p_2$ and $\tilde{p}_2$ and the boundary point $b_u^R$ of the gravitational region. 
 \item \underline{Group 3}: $n_1$ number of $w$-planes with a cut connecting two bulk points of the gravitational region denoted as $p_3$ and $\tilde p_3$ and the boundary point $b_u^L$ of the gravitational region. 
 \item \underline{Group 4}: $n_1$ number of $w$-planes with a cut connecting two bulk points of the gravitational region denoted as $p_4$ and $\tilde p_4$ and the boundary point $b_u^R$ of the gravitational region. 
 \item \underline{Group 5}: $n_2$ number of $w$-planes with a cut connecting two bulk points of the gravitational region denoted as $p_5$ and $\tilde p_5$ and the boundary point $b_u^L$ of the gravitational region. 
 \item \underline{Group 6}: $n_2$ number of $w$-planes with a cut connecting two bulk points of the gravitational region denoted as $p_6$ and $\tilde p_6$ and the boundary point $b_u^L$ of the gravitational region. 
 \item \underline{Group 7}: $n_2$ number of $w$-planes with a cut connecting two bulk points of the gravitational region denoted as $p_7$ and $\tilde p_7$ and the boundary point $b_u^L$ of the gravitational region. 
 \item \underline{Group 8}: $n_2$ number of $w$-planes with a cut connecting two bulk points of the gravitational region denoted as $p_8$ and $\tilde p_8$ and the boundary point $b_u^L$ of the gravitational region. 
 \item \underline{Group 9}: $w$-plane having eight cuts. The $i^{\text{th}}$ cut connecting two bulk points $q_i, \tilde q_i$ and the boundary point $b_u^L$ or $b^R_u$ as shown in figure \ref{fig:USinpathintegralwormholeswmap13}. 
 \end{itemize}
The points $p_1,p_2,p_3$ and $p_4$  have deficit angle of $2\pi-2\alpha_1$ and the points $p_5,p_6,p_7$ and $p_8$  have deficit angle of $2\pi-2\alpha_2$. The end points of the cuts $q_i, ~i=1,2,3,4$ have deficit angle  $2\pi-2\zeta_1$ and the deficit angle at points $q_i, ~i=5,6,7,8$ is $2\pi-2\zeta_2$. \par

Then the surface $\tilde{\mathcal{G}}_{n,0}$ is obtained by the cyclic gluing of the cut interval $\left(\tilde{p}_i, b_u^R\right)$ of the $w$-planes in group $i$ and the cut interval $\left( \tilde q_i,b_u^R\right)$ in the $w$-plane of group 9 for $i=1,4,5,8$ and the cyclic gluing of the cut interval $\left(\tilde{p}_j, b_u^L\right)$ of the $w$-planes in group j and the cut interval $\left( \tilde q_j,b_u^L\right)$ in the $w$-plane of group 9 for $i=2,3,6,7$. \par

The partition function of the matter CFT on the wormhole ${\mathcal{G}}_{n,0}$ can be obtained by gluing the partition function on the cut geometry $\tilde{\mathcal{G}}_{n,0}$ appropriately along its boundaries. As described above, $\tilde{\mathcal{G}}_{n,0}$ is a hyperbolic Riemann surface with several points having deficit and excess angles. Therefore the universal terms in the free energy of the wormhole ${\mathcal{G}}_{n,0}$ can be identified as the JT gravity and CFT contributions coming from the points on $\tilde{\mathcal{G}}_{n,0}$ having deficit and excess angles. Instead of searching for a saddle in the full space of wormhole geometries, we will restrict our search with the space of the special class of wormhole geometries having the property that the points $q_1,q_2,q_3$ and $q_4$ on them are very close to the points $q_8,q_7,q_6$ and $q_5$ respectively. Similarly, the points $p_1,p_2,p_3$ and $p_4$ are assumed to be very close to the points $p_8,p_7,p_6$ and $p_5$ respectively.  Such a wormhole geometry can be constructed by gluing $n$ number of $w$-planes as shown in figure \ref{fig:USinpathintegralW}. The universal terms in the free energy of such a wormhole geometry is given by
\begin{align}\label{FW}
F_W&\approx-\frac{2\pi-2\zeta_1+2\pi-2\zeta_2}{4\pi G_N}\sum_{i=1}^4\Phi\left(w_{q_i},\bar w_{q_i}\right)-\frac{n_1(2\pi-2\alpha_1)+n_2(2\pi-2\alpha_2)}{4 \pi G_N}\sum_{i=1}^4\Phi\left(w_{p_i},\bar w_{p_i}\right)\nonumber\\
&+\frac{c}{24}\left[n_1\Theta\left(2\pi -2\alpha_1\right)+n_2\Theta\left(2\pi -2\alpha_2\right)\right]\text{ln}~\left(w_{p_1}-w_{b_u^R}\right)\left(w_{p_2}-w_{b_u^L}\right)\left(w_{p_3}-w_{b_u^L}\right)\left(w_{p_4}-w_{b_u^R}\right)\nonumber\\
&+\frac{c}{24}\left[\sum_{i=1}^2\left(\Theta\left(2\pi n_i\right) +\Theta\left(2\pi -2\zeta_i\right)\right)\right]\text{ln}~\left(w_{q_1}-w_{b_u^R}\right)\left(w_{q_2}-w_{b_u^L}\right)\left(w_{q_3}-w_{b_u^L}\right)\left(w_{q_4}-w_{b_u^R}\right)
\nonumber\\
&+c.c.
\end{align}
The end points of the cuts $p_i$ and $q_i$ for $ i=1,\cdots,4$ can be determined by minimising $F_W$.  It is straightforward to verify that the free energy of the resulting wormhole saddle after the Lorentzian continuation is time independent. We shall denote the resulting wormhole saddle as ${\mathcal{G}}^W_{n,0}$. 

\subsection{The action of the reduced half-sided translations after Page time}
 In this subsection, we will describe the action of the reduced half-sided translations after the Page time by analysing the CFT path integral on the wormhole saddle ${\mathcal{G}}^W_{n,0}$. We do this by interpreting the  CFT path integral on ${\mathcal{G}}^W_{n,0}$ as a result of the back-reaction of the modified reduced half-sided translation operator $U^{E_u}_{W}(s)$ placed on the $w$-plane along appropriate cuts. The modified cuts can be easily identified if the operator $U^{E_u}_{W}(s)$ is close to identity, since the backreaction is negligible in this case. This is the case if $s$ infinitesimal.  Therefore, we will consider ${\mathcal{G}}^W_{4n_1+4n_2+1,0}$ for $$n_2=-n_1=\epsilon,$$ where $\epsilon$ is an infinitesimal parameter.  For this saddle, the associated angles $\alpha_i$ and $\zeta_i$ satisfying \eqref{csingremoval} can be approximated as
 \begin{align}\label{alphaizetai}
\alpha_i=\pi-(-1)^i\epsilon \pi \qquad \zeta_i=\pi-(-1)^i\epsilon \pi+\epsilon^2\pi \qquad i=1,2.
\end{align}
  The free energy of this wormhole saddle is obtained by substituting \ref{alphaizetai} in \ref{FW} 
 \begin{align}
F_W&\approx  \frac{\epsilon^2}{ G_N}\sum_{i=1}^4\Phi\left(w_{q_i},\bar w_{q_i}\right)-\frac{\epsilon^2}{ G_N}\sum_{i=1}^4\Phi\left(w_{p_i},\bar w_{p_i}\right)\nonumber\\
&+\frac{c\epsilon^2}{6}\text{ln}~\left(w_{p_1}-w_{b_u^R}\right)\left(w_{p_2}-w_{b_u^L}\right)\left(w_{p_3}-w_{b_u^L}\right)\left(w_{p_4}-w_{b_u^R}\right)\nonumber\\
&-\frac{c\epsilon^2}{6}\text{ln}~\left(w_{q_1}-w_{b_u^R}\right)\left(w_{q_2}-w_{b_u^L}\right)\left(w_{q_3}-w_{b_u^L}\right)\left(w_{q_4}-w_{b_u^R}\right)
\nonumber\\
&+c.c.
\end{align}
Then the cuts can be determined by identifying the end points of the cuts by minimising the free energy.  Minimisation gives that  $p_1=q_1=p_4=q_4$ and $p_2=q_2=p_3=q_3$.\par

Let us denote the density matrix associated with the $w$-plane having a cut connecting $p_1$ and $b_u^R$ be $\rho_R$, and the density matrix associated with the $w$-plane having a cut connecting $p_2$ and $b_u^L$ be $\rho_L$. Then the path integral on the wormhole saddle ${\mathcal{G}}^W_{n,0}$ can be interpreted as the correlation function 
\begin{align}
\langle \Omega|\cdots \rho_R^{n_1}\rho_L^{n_2}\rho_R^{n_2}\rho_L^{n_1}\mathcal{O}\rho_L^{n_2}\rho_R^{n_1}\rho_L^{n_1}\rho_R^{n_2} \cdots |\Omega\rangle \qquad n_2=-n_1=\text{i}t.
\end{align}
 Clearly, the reduced half-sided translation $U^{E_u}_W(s)$  can be identified with the identity operator. This implies that the action of the reduced half-sided translation constructed using the degrees of freedom available inside the gravitational region outside the black hole horizon is trivial at late times. Consequently, after the Page time the local operators inside the horizon of the black hole can not be reconstructed using the degrees of freedom available inside the gravitational region outside the black hole horizon. This is consistent with the fact that after the Page time, the black hole interior is contained in the island region of the bath entanglement wedge \cite{Almheiri:2019yqk}.
 
\section{Discussion}\label{disc}

In this paper, we showed that after Page time, the half-sided modular translations defined using the algebra of observables restricted to the gravitational region of black hole in equilibrium with a bath become trivial. The bulk half-sided modular translations can also be interpreted as appropriate boundary half-sided modular translations \cite{Leutheusser:2022bgi}. Therefore, our result can be interpreted as the inability of the Liu-Leutheusser proposal to reconstruct the interior bulk operators using the boundary CFTs, the CFT$_R$ and the CFT$_L$ after Page time. This implies, after Page time, the black hole interior does not belong to the entanglement wedge of the dual boundary CFTs \cite{Almheiri:2019yqk}. Thus the result upholds the general principles of the entanglement wedge reconstruction \cite{Dong:2016eik}. \par

Let us briefly mention some of the directions that deserve immediate attention. It is known that non-trivial half-sided translation exists only if the von Neumann operator algebra is type III$_1$ \cite{Wiesbrock:1992mg}. Hence, we expect that the algebra of observables restricted to the gravitational region outside the horizon of the black hole in equilibrium with a bath have type III$_1$ algebra before Page time. However, our result suggests that after Page time the algebra of observables must cease to be a type III$_1$ algebra. This can be checked by finding the spectrum of the logarithm of the modular operator of the operator algebra. If the spectrum does not cover the full real axis, then the operator algebra of our interest is not of type III$_1$ \cite{connes1973classification,connes1974caracterisation}. Another interesting direction is to check that after Page time whether half-sided translation allows us to express a field inside the black hole interior in terms of degrees of freedom of bath\footnote{This is already done in \cite{Jalan:2024cby}}. 

\section*{Acknowledgements}
We would like to thank Debodirna Ghosh for the initial collaboration and the discussions. We are indebted to Sujay Ashok and Alok Laddha for the numerous illuminating discussions and the comments. We also thank Abhijit Gadde, Hao Geng, Shiraz Minwalla, Sitender Kashyap, Venkata Suryanarayana Nemani, Suvrat Raju, Manish Ramchander, Pratik Rath, Ashoke Sen, Sandip Trivedi and Amitabh Virmani for the discussions.

\bibliographystyle{JHEP}
\bibliography{reference}

\end{document}